# Application of Group Method of Data Handling and New Optimization Algorithms for Predicting Sediment Transport Rate under Vegetation Cover

Golnaz Mirzakhani[a], Elham Ghanbari-Adivi[a]*, Rohollah Fattahi[a], Mohammad Ehteram[b], Amir Mosavi[c,d,e]* Ali Najah Ahmed[f], Ahmed El-Shafie[g,h]

a) Department of Water Science Engineering, Shahrekord University, Shahrekord, Iran
b) Department of Water Engineering and Hydraulic Structures, Faculty of Civil Engineering, Semnan University, Semnan, Iran
c) John von Neumann Faculty of Informatics, Obuda University, 1034 Budapest, Hungary
d) German Research Center for Artificial Intelligence (DFKI), Kaiserslautern, Rheinland-Pfalz, Germany
e) Institute of Information Engineering, Automation and Mathematics, Slovak University of Technology in Bratislava, Bratislava, Slovakia
f) Department of Civil Engineering, College of Engineering, Universiti Tenaga Nasional (UNITEN), 43000 Selangor, Malaysia
g) National Water and Energy Center, United Arab Emirates University, P.O. Box. 15551, Al Ain, United Arab Emirates
h) Department of Civil Engineering, Faculty of Engineering, University of Malaya (UM), 50603 Kuala Lumpur, Malaysia
* Corresponding author Email: ghanbariadivi@sku.ac.ir, amir.mosavi@kvk.uni-obuda.hu

## Abstract

Planting vegetation is one of the practical solutions for reducing sediment transfer rates. Increasing vegetation cover decreases environmental pollution and sediment transport rate (STR). Since sediments and vegetation interact complexly, predicting sediment transport rates is challenging. This study aims to predict sediment transport rate under vegetation cover using new and optimized versions of the group method of data handling (GMDH). Additionally, this study introduces a new ensemble model for predicting sediment transport rates. Model inputs include wave height, wave velocity, density cover, wave force, D50, the height of vegetation cover, and cover stem diameter. A standalone GMDH model and optimized GMDH models, including GMDH- honey badger algorithm (HBA), GMDH- rat swarm algorithm (RSOA), GMDH- sine cosine algorithm (SCA), and GMDH- particle swarm optimization (GMDH-PSO), were used to predict sediment transport rates. As the next step, the outputs of standalone and optimized GMDH were used to construct an ensemble model. The MAE of the ensemble model was 0.145 m3/s, while the MAEs of GMDH-HBA, GMDH-RSOA, GMDH-SCA, GMDH-PSOA, and GMDH in the testing level were 0.176 m3/s, 0.312 m3/s, 0.367 m3/s, 0.498 m3/s, and 0.612 m3/s, respectively. The Nash–Sutcliffe coefficient (NSE) of ensemble model, GMDH-HBA, GMDH-RSOA, GMDH-SCA, GMDH-PSOA, and GHMDH were 0.95 0.93, 0.89, 0.86, 0.82, and 0.76, respectively. Additionally, this study demonstrated that vegetation cover decreased sediment transport rate by 90%. The results indicated that the ensemble and GMDH-HBA models could accurately predict sediment transport rates. Based on the results of this study, sediment transport rate can be monitored using the IMM and GMDH-HBA. These results are useful for managing and planning water resources in large basins.

## 1. Introduction

Pollution of the environment may result from sediment transport. Furthermore, sediment transportation may reduce the capacity of dam reservoirs. Coastal residents and downstream hydraulic structures are adversely affected by sediment transport (Jalil-Masir et al., 2021). Thus, watershed management requires predicting sediment transport rate (STR) (Jalil-Masir et al., 2021). Sediment is efficiently trapped by vegetation cover. Sediment traps can protect rivers from suspended loads (Jalil Masir et al., 2022). Sediment transport rate refers to the amount of sediment that passes through a given flow-transverse cross section per unit of time. The STR is commonly expressed in volume terms.

It is challenging for managers and decision-makers to predict sediment transport rate (STR) under vegetation cover (Fathi-Moghadam et al., 2018). Since sediment and vegetation cover interact complexly, predicting sediment transport rates in the presence of vegetation cover can be difficult. Many studies have been conducted to predict sediment transport rate (STR) in the presence of vegetation coer. The effect of vegetation density, water depth, and sediment grain size on sediment transport was studied by Wang et al. (2015). They reported that vegetation decreased sediment transport. Based on Igarashi and Tanaka's (2016) report, integrating coastal forests and embankments could reduce wave force by 80 percent. Chen et al. (2018) examined the sediment transport rate in a bare mudflat and a mangrove stand and reported that vegetation altered the flow rate. Permatasari et al. (2018) investigated the correlation between mangrove density and sediment transport. There was a negative correlation between mangrove density and sediment transport. Parnak et al. (2018) investigated the effect of rigid and flexible vegetation covers on the sediment transport rate. They reported that vegetation cover could significantly reduce the sediment transport rate by 70%. Mu et al. (2019) found that basal stem covers drastically reduced the transport capacity of overland flows. Kusumoto et al. (2020) reported that coastal forest cover reduced sediment transport. Sun et al. (2020) stated that ecological restoration affected sedimentary delivery.

Planting vegetation to trap sediment is a complex issue. Therefore, hydraulic engineers are responsible for forecasting sediment transport in vegetation-covered areas (Jalil-Masir et al., 2021). However, this prediction requires powerful models. Using numerical models to predict sediment transport rate (SRT) is possible, but they require solving complicated equations (Jalil-Masir et al., 2021). STR can also be predicted using empirical equations, but they are not accurate enough. The empirical equations are based on the experimental data. The use of soft computing models to predict sediment transfer rates (STR) has become increasingly popular in recent years. The soft computing models include artificial intelligence, support vector machines, decision tree, and adaptive fuzzy interface system models. The soft computing models include artificial neural network models, adaptive neuro-fuzzy interface systems, support vector machines, and group methods of data handling Table 1 reviews soft computing models' application to sediment transfer rate (STR) prediction. Since soft computing models are accurate, fast, and easy to use, they can efficiently predict sediment transfer rates (Liang et al., 2021).

Table 1. Application of soft computing models for predicting STR

| Authors | Description | Results |
| --- | --- | --- |
| Ab. Ghani and Azamathulla (2014) | They evaluated gene expression programming (GEP) for estimating STR. | The suggested GEP model gave accurate results for predicting STR |
| Kitsikoudis et al. (2015) | They applied an adaptive network-based **fuzzy** inference system (ANFIS) for STR prediction. | They reported that the ANFS model provided better accuracy than *the* empirical methods. |
| Ebtehaj and Bonakdari (2016a) | They tested several models for the estimation of STR. They compared different training algorithms for estimating STR. | They found that the artificial neural network- Levenberg-Marquardt performed better than existing equations. |
| Ebtehaj and Bonakdari (2016b) | They compared two soft computing models for STR prediction. | It was found that the extreme learning machine model performed better than the support vector machine model. |
| Roushanger and Ghasempour (2017) | They estimated the STR in pipes using a support vector machine (SVM). | It was found that the SVM method was superior to classical methods. |

| Riahi-Madvar and Seifi (2018) | They estimated STR in gravel-bed rivers using two soft computing models. | It was found that the ANFIS model was superior to the ANN model. |
| Baniya et al. (2019) | They estimated STR using the bed shear stress ($\tau_b$), specific stream power ($\omega$), and flow velocity (v). They investigated the potential of the ANN model for predicting STR. | It was found that the ANN model was superior to all the suggested models. |
| Kargar et al. (2019) | They compared two soft computing models for STR prediction. | They found that the neuro-fuzzy model performed better than the genetic programming model. |

Previous studies used soft computing models to predict sediment transfer rates (STRs) without considering cover vegetation. Furthermore, the use of individual models is one of the shortcomings of the former studies. While the individual model may compute quickly, it is not highly accurate. Ensemble models based on individual model outputs may improve final output accuracy (Panahi et al., 2021). Watershed managers can use soft computing models to predict sediment transfer rates (STRs) in vegetation-covered watersheds. Table 1 shows that the artificial neural network is one of the feasible models for predicting sediment transfer rates (STRs). There are several types of artificial neural networks. One of the most important kinds of ANN models is the group method of data handling (GMDH). The GMDH model is based on the principle of self-organization. Polynomial transfer functions and multiple neuronal layers are used in the GMDH algorithm. High speed and accuracy are two advantages of GMDH. There are a variety of applications of the GMDH, including flood susceptibility prediction (Dodangeh et al., 2020), groundwater level prediction (Moosavi et al., 2020), daily river flow prediction (Aghelpour et al.,

2020), and monthly streamflow prediction (Adnan et al., 2021). Mulashani et al. (2022) used GMDH to predict permeability. With the GMDH, processing time was reduced, and accuracy was increased. Landslides were spatially modeled using the GMDH model by Panahi et al. (2022). Validation results showed that optimized GMDH models performed better than standalone GMDH models. In addition to its high accuracy, GMDH models are easy to implement and quick to compute. GMDH is a robust model, but robust training algorithms are necessary to determine its weight coefficients. Robust optimization algorithms can be used for adjusting GMDH parameters.

Previous studies have used soft computing models to predict STR, but there are still several gaps. Previous studies only used individual soft computing and optimized models to predict STR. Literature reviews indicated that ensemble models outperformed individual soft computing models. STR was not predicted using ensemble models. In previous studies, STR was estimated through experimental studies without considering robust models. So, it is imperative to develop soft computing models to predict STR. This study simulates the effect of vegetation cover on the STR, where several studies ignore it. A new ensemble model is also introduced to fill previous research gaps in this study.

This study investigates the effect of vegetation on sediment transport rate and presents the details of a comprehensive experiment. By using the new optimization algorithm, GMDH models are developed to predict sediment transport rates under cover vegetation. This study examines how vegetation affects sediment transport rates by predicting sediment transport rates using a variety of input parameters. This study also presents an ensemble model to assemble the outputs of GMDH models, which can be used to improve the accuracy of individual models. An ensemble framework increases the efficiency of each individual model by combining the outputs of multiple models.

Therefore, the current article contains the following novelties:

1- Using new optimization algorithms, novel GMDH models predict sediment transport rates. For training the GMDH model, new optimization algorithms were used. The high accuracy, fast computation, and flexibility of algorithms make them the ideal choice.

2. The inclusive multiple model (IMM) is proposed as a new ensemble model for predicting sediment rate.
3. The effect of cover vegetation on sediment transport is investigated experimentally.
4. The effect of various cover vegetation layouts on the sediment transport rate is determined.
5. Predicting sediment transport rates is examined using various inputs.

The material and methods are presented in Section 2. Experiment details are presented in Section 3. The discussion and results are presented in Section 4. Lastly, Section 5 concludes the paper.

## 6. Materials and Methods

Different optimization algorithms have been developed to solve problems in recent years. Wang et al. (2019) introduced a new optimization algorithm, the monarch butterfly optimization algorithm (MBOA). The migration of monarch butterflies was used to develop the monarch butterfly optimization algorithm. (MBOA). Li et al. (2020) proposed the slime mould algorithm (SMA) as an advanced stochastic optimizer. The SMA simulates slime mould's search for food by adding weight. A new metaheuristic algorithm, called Moth Search (MS), was presented by Wang et al. (2018). Hunger Games Search (HGS) is a technique developed by Yang et al. (2021). HGS is based on the behavioral choices and hunger-driven activities of animals. A Runge Kutta optimizer (RUN) was developed by Ahmadifar et al. (2018). The RUN method is a promising and logical way to search for global optimization. This study presented several algorithms for training

GMDH, including the Honey Badger Algorithm, Rat Swarm Optimization Algorithm, Sine Cosine Optimization Algorithm, and Particle Swarm Optimization Algorithm (PSOA). These algorithms have fast convergence, high accuracy, and ease of implementation. This study chose these algorithms because of these reasons.

## 2.1 Structure of group method of data handling

GMDH comprises an input, hidden, and output layers (Radiadeh and Kozlowski, 2020). Input variables are inserted into the input layer. A hidden layer of GMDH processes the data received from the previous layer
. Lastly, the output layer produces the desired result. The GMDH constructs a high-order polynomial named Kolmogorov-Gabor as follows

$$output = \alpha_0 + \sum_{i=1}^{d} \alpha_i in_i + \sum_{i=1}^{d}\sum_{j=1}^{d} \alpha_{ij} in_i in_j + \sum_{i=1}^{d}\sum_{j=1}^{d}\sum_{k=1}^{d} \alpha_{ijk} in_i in_j in_k \tag{1}$$

Where $output$: final output, $\alpha_0$, $\alpha_i$, $\alpha_{ijk}$, and $\alpha_{ij}$: polynomial coefficients, $in_i, in_j, in_k$: ith, jth, and kth input and d: number of inputs. In this research, the quadratic form of the polynomial was utilized. For example, the quadratic form for a problem with two inputs is as follows (Radaideh and Kozlowski, 2020):

$$output = \alpha_0 + \alpha_1 in_1 + \alpha_2 in_2 + \alpha_3 in_3^2 + \alpha_4 in_2^2 + \alpha_5 in_1 in_2 \tag{2}$$

The polynomial coefficient vector is computed as follows:

$$\alpha = \left(\beta^T \beta\right)^{-1} \beta^T OUT \tag{3a}$$

Where $\beta$: a matrix based on inputs and T: transpose.

$$\beta = \begin{bmatrix} 1 & in_1^1 & in_2^1 & in_1^1 in_2^1 & \left(in_1^1\right)^2 & \left(in_2^1\right)^2 \\ . & in_1^2 & in_2^2 & in_1^2 in_2^2 & \left(in_1^2\right)^2 & \left(in_2^2\right)^2 \\ . & . & . & . & . & . \\ . & . & . & . & . & . \\ 1 & in_1^m & in_2^m & in_1^m in_2^m & \left(in_1^m\right)^2 & \left(in_2^m\right)^2 \end{bmatrix} \quad (3b)$$

Where $OUT$: matrix of outputs. The following equation gives the number of neurons in the next layer ($N_n$):

$$N_n = \binom{N_{np}}{2} \quad (4)$$

Where $N_{np}$: number of the current layers. The number of neurons in the layers is limited to a maximum value to prevent the complexity of GMDH. Users can use an equation to eliminate GMDH's redundant neurons. The neurons with lower RMSE than the $\xi$ (selection-pressure criterion) are removed.

$$\xi = \upsilon \times RMSE_{\min} + (1-\upsilon) \times RMSE_{\max} \quad (5)$$

Where $\upsilon$: a value between 0 and 1, $RMSE_{\max}$: the RMSE of the worst neuron, $RMSE_{\min}$: the RMSE of the best neuron. Polynomial coefficients can be adjusted using the backpropagation algorithm (BPA), although it may not have a rapid convergence. In this research, robust evolutionary algorithms were used to set the polynomial coefficients of the GMDH. Figure 1a depicts the structure of GMDH (Radiadeh and Kozlowski, 2020).

### 2.2 Honey Badger Algorithm (HBA)

Hashim et al. (2022) introduced the HBA algorithm as a novel optimization technique. A variety of engineering problems were used to evaluate HBA. HBA has the advantages of rapid convergence, high precision, and high diversity. Due to these advantages, the HBA was chosen for the current study. Honey badgers are attracted to honey and build holes to trap food. Badgers stay in holes to mate.Badgers stay in holes to mate. For access to beehives and nests, they climb trees. Locating beehives is one of honey badgers' challenges. When honey badgers locate beehives, they are assisted by honeyguides (birds). HBs can hunt squirrels and lizards and use their sense of smell to locate prey. In the first level, the location of HBs is initialized as follows (Hashim et al., 2022):

$$HB_i = lo_i + ra_1 \times (up_i - lo_i) \tag{6}$$

Where $HB_i$: the location of HBs, $up_i$: the upper bound of decision variable, $lo_i$: lower bound of decision variable, and $ra_1$: random variable. The HBs can identify the location of prey or honey based on the intensity of the smell they receive.

$$In_i = ra_2 \times \frac{S}{4\pi di_i^2} \tag{7}$$

$$S = (HB_i - HB_{i+1}) \tag{8}$$

$$di_i = HB_{prey} - HB_i \tag{9}$$

Where $HB_i$: the location of ith HB, $HB_{i+1}$: the location of i+1th HB, $ra_2$; random number, and $In_i$: smell intensity, $S$: source strength, and $HB_{prey}$: location of prey, and $di_i$: distance between rye and HB. The HBs excavate holes to rest and trap prey. The location of HB following digging is calculated as follows:

$$HB_{new} = HB_{prey} + F \times \rho \times In_i \times HB_{prey} + F \times ra_3 \times \eta \times di_i \times |\cos(2\pi ra_4) \times [1 - \cos(2\pi ra_5)]| \quad (10)$$

Where $HB_{prey}$: prey location, F: flag (it changes direction), $ra_3$ and $r_{a4}$: random parameters, $\rho$: controller parameter, and $\eta$: density factor. The density factor is computed as follows:

$$\eta = Z \times \exp\left(\frac{-t}{t_{max}}\right), t_{max} = \max imum(number) of (iterations) \quad (11)$$

$$F = \begin{bmatrix} 1 \leftarrow if(r_6) \leq 0.50 \\ -1 \leftarrow else \end{bmatrix} \quad (12)$$

Where t: number of iterations and $r_6$: random number.

Additionally, HBs update their location when they follow honeyguide birds to reach beehives.

$$HB_{new} = HB_{prey} + F \times ra_7 \times \eta \times di_i \quad (13)$$

Where $HB_{new}$: new location of the HB after following honeyguide bird and $ra_7$: random parameter. Figure 1b depicts the flowchart of the HB.

## 2.3 Rat Swarm Optimization algorithm (RSOA)

Dhiman et al. (2021) developed the RSOA based on rat behavior. ROSA's advantages include ease of execution and a small number of random parameters. They evaluated RSOA's performance on various benchmark functions and engineering problems. The ROSA algorithm outperformed the particle swarm optimization (PSO), the genetic algorithm (GA), the gravitational search algorithm

(GSA), and the multiverse optimization algorithm (MOA). Rats exhibit aggressive behavior and are constantly in search of prey. A swarm of rats will follow the superior rat, aware of its prey's location. As a result, the rats update their location following the location of the rat leader as follows:

$$\vec{RA} = C.\vec{RA_i}(x) + S.(\vec{RA_r}(x) - \vec{RA_i}(x)) \tag{14}$$

Where $\vec{RA}$: the new location of the rat, $\vec{RA_i}(x)$: the ith location of the rat, $\vec{RA_r}(x)$: the best location of the rat, $S$ and C: controller parameters.

$$C = \gamma - x \times \frac{\gamma}{\max_{iter}}, x = 0,..,\max_{iter} \tag{15}$$

$$S = 2.rand \tag{16}$$

Where $\gamma$: a number between [1, 5], rand: random number, and max$_{iter}$: maximum number of iterations. Rats struggle with prey to hunt it. This behavior is simulated as follows:

$$\vec{RA_i}(x+1) = |\vec{RA_r}(x) - \vec{RA}| \tag{17}$$

Where $\vec{RA_i}(x+1)$: The subsequent position of the rat. Figure 1c illustrates the RSOA flowchart.

### 2.4 Structure of Sine Cosine Algorithm

Based on sine and cosine trigonometric functions, the Sine Cosine algorithm (SCA) is a novel optimization technique (Mirjalili et al., 2016). SCAs are robust, adaptable, and have a high convergence rate (Abualigah and Diabat, 2021). A population's solution vector represents a candidate solution. The optimal solution is selected as the final destination. The SCA algorithm updates the solution based on the following equation:

$$S_i^{t+1} = \begin{bmatrix} S_i^t + \kappa_1 \times \sin(\kappa_2) \times |\kappa_3 Z_i^t - S_i^t| \leftarrow \kappa_4 < 0.50 \\ S_i^t + \kappa_1 \times \cos(\kappa_2) \times |\kappa_3 Z_i^t - S_i^t| \leftarrow \kappa_4 \geq 0.50 \end{bmatrix} \quad (18)$$

Where $\kappa_1$: A parameter for controlling the balance between exploration and exploitation $\kappa_2$: A parameter for determining the direction of solutions, $\kappa_3$: A parameter for adjusting stochastic influence of the global best solution, and $\kappa_4$: A parameter for determining the priority of sine function or cosine function.

$$\kappa_1 = \psi - \frac{\psi i}{I} \quad (19)$$

Where $\psi$: constant value, i: number of iterations, I: maximum number of iterations. Figure 1d shows the SCA flowchart.

### 2.5 Particle Swarm Optimization Algorithm (PSOA)

The PSO algorithm is known for its robustness and its ability to exchange information between particles. Achieving a good balance between exploration and exploitation is one of the advantages of PSO (Achite et al., 2022). A PSO begins by determining the particle's position and velocity (Ehteram et al., 2021). A PSO begins by determining the particle's position and velocity (Ehteram et al., 2021). Based on the optimal objective function, an optimal particle was identified. The velocity and location of particles are updated based on the following equation:

$$ve_i^{t+1} = \varsigma ve_i^t + \mu_1 (P_g - p_i^t) + \mu_2 (P_i^* - p_i^t) \quad (20)$$

$$p_i^{t+1} = p_i^t + v_i^{t+1} \tag{21}$$

Where $ve_i^{t+1}$: velocity of i+1th at t+1th iteration, $ve_i^t$: velocity of ith at iteration t, $P_g$: current global best solution, $P_i^*$: individually best solution, $\mu_1$ and $\mu_2$: acceleration coefficient, $p_i^t$: the location of the ith particle, $\varsigma$: inertia coefficient, and $p_i^{t+1}$: the location of a particle at t+1 iteration (Figure 1f).

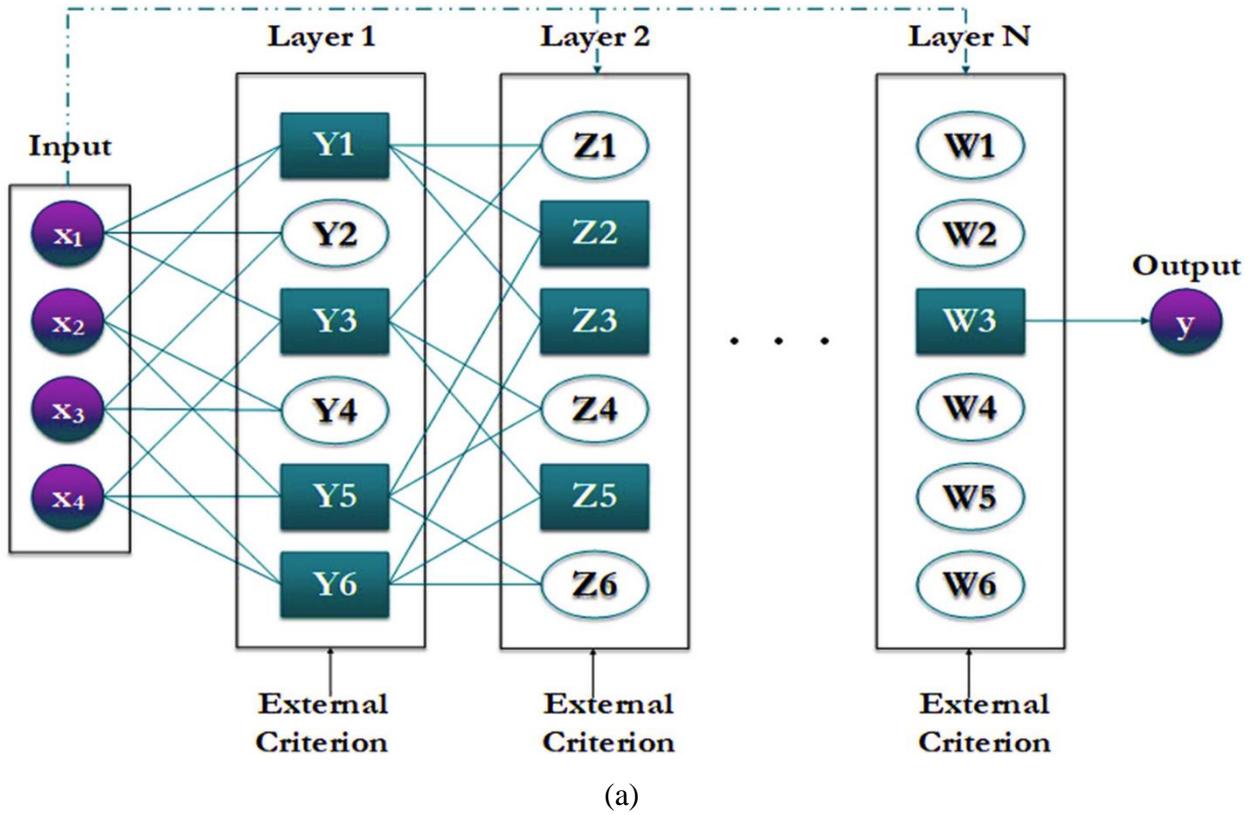

(a)

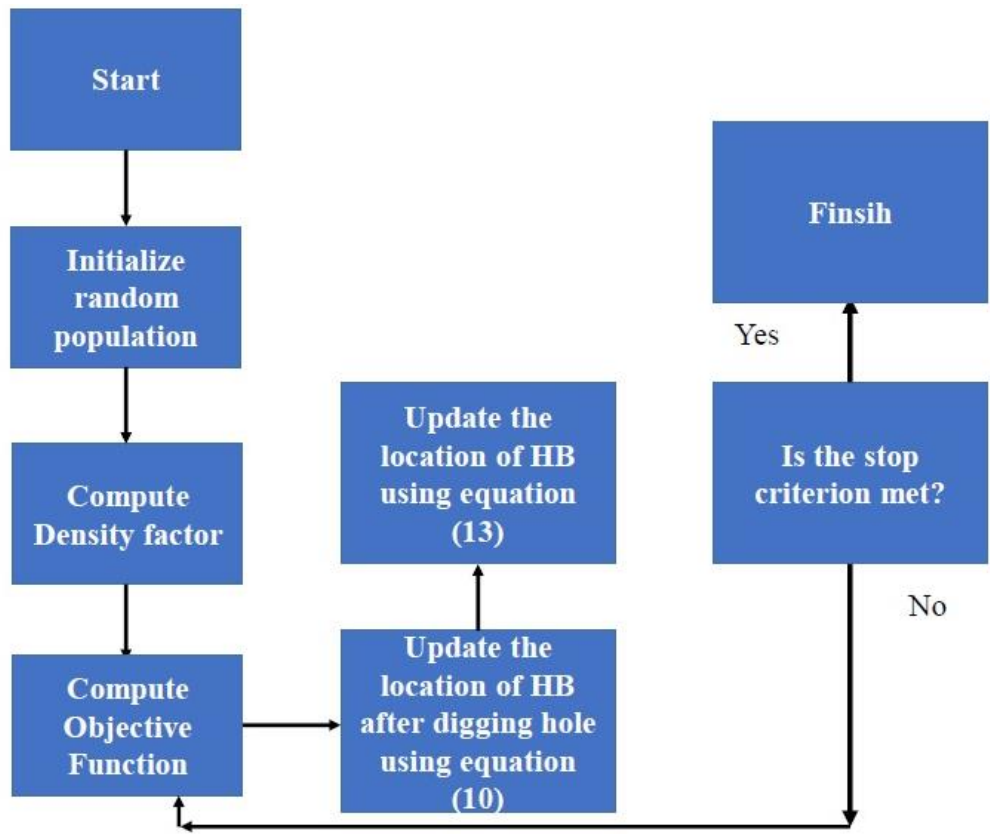

b

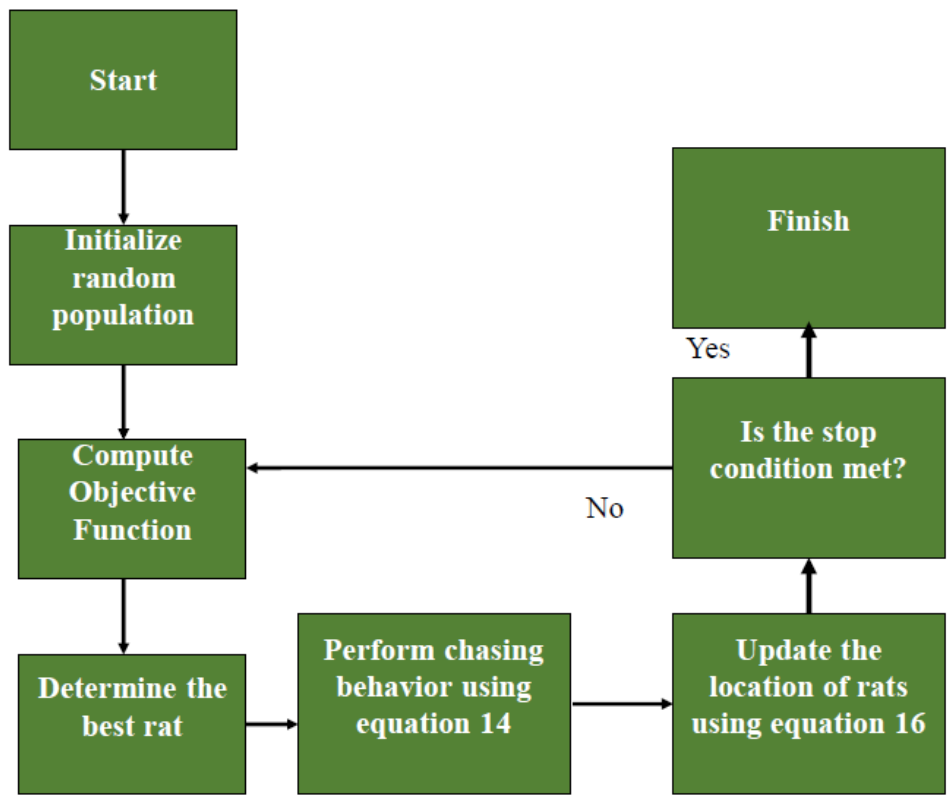

c

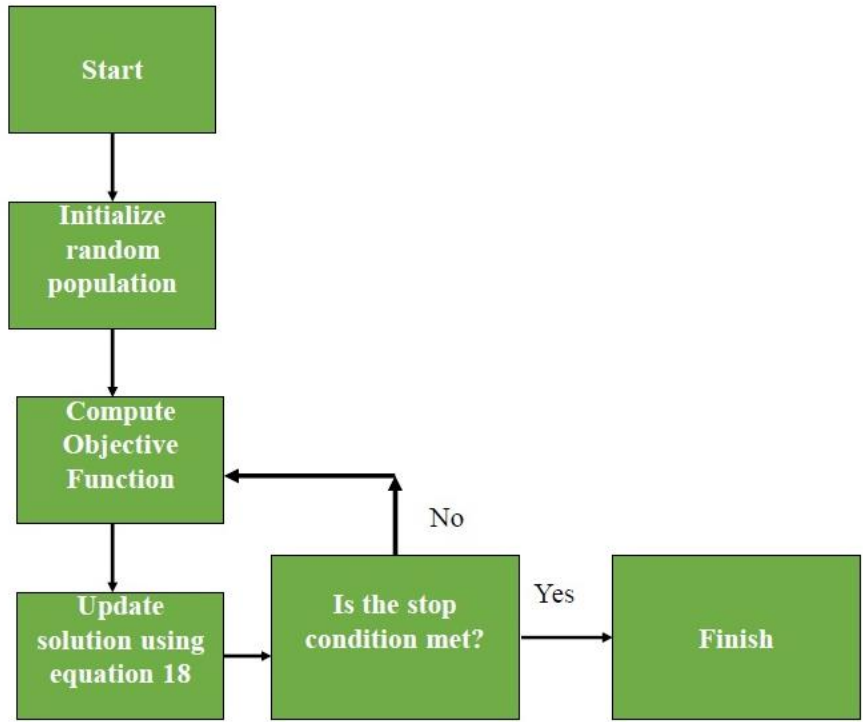

d

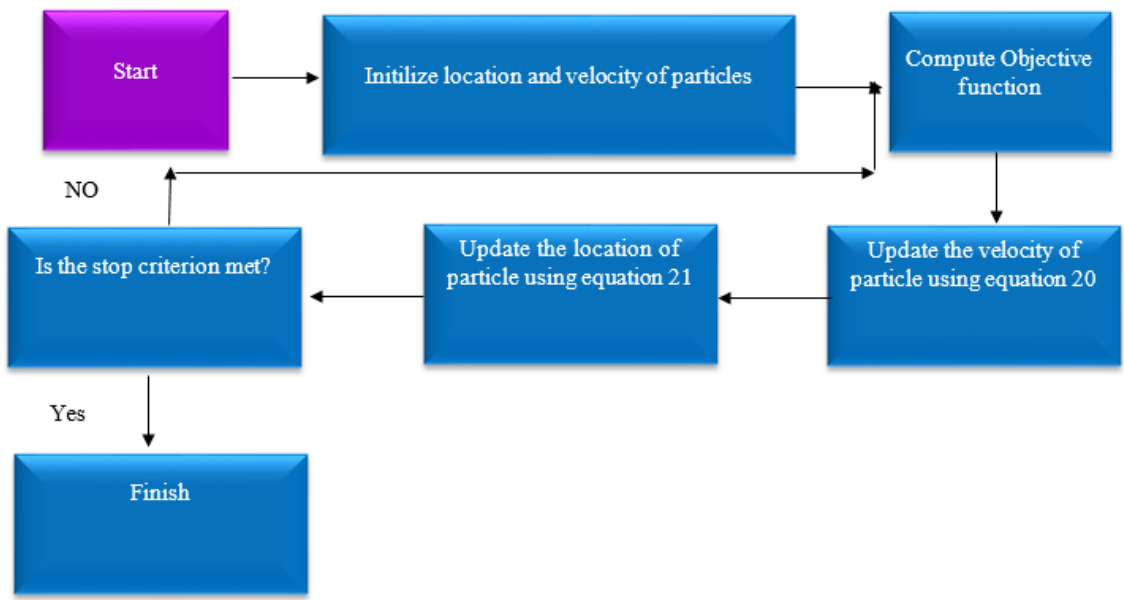

e

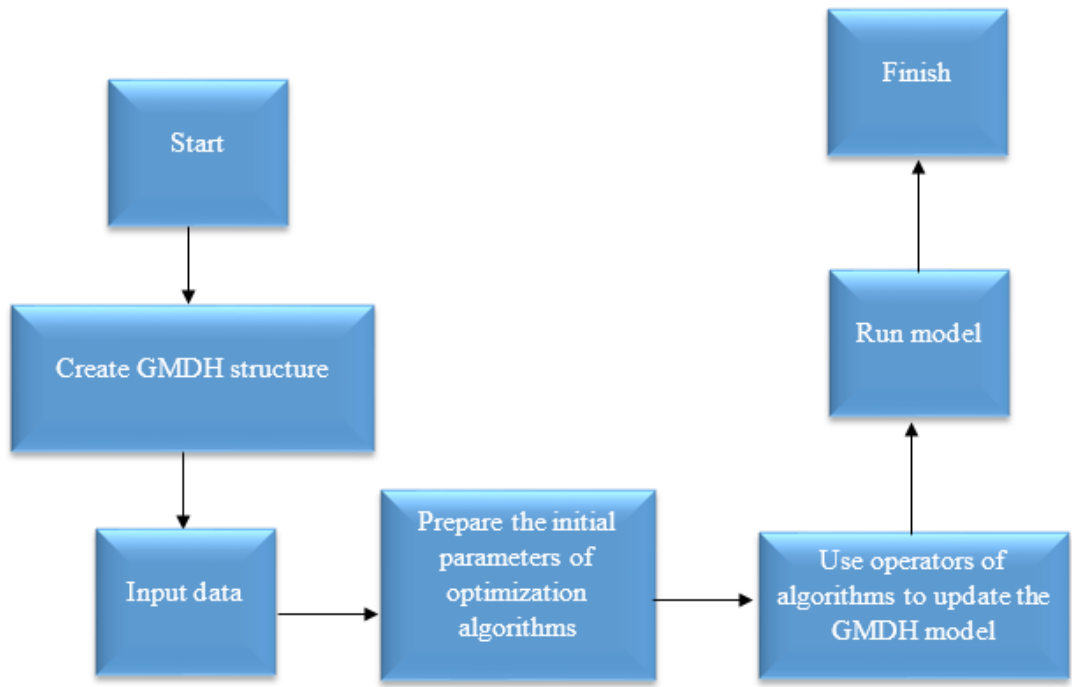

f

Figure 1. a: Structure of GMDH (Radiadeh and Kozlowski, 2020), b: The flowchart of HB, c: The Flowchart of RSOA, d: The SCA flowchart, e: PSO flowchart  f:The structure of optimized GMDH model

**2.6 GMDH integration with optimization algorithms**

1- The data are divided into training and testing data.

2- The initial values of polynomial coefficients were inserted into the GMDH.

3- The GMDH model is run at the training level.

4- If the stop condition is met, the model is run at the testing level; otherwise, the optimization algorithms are used to adjust the GMDH parameters.

5- The initial values of polynomial coefficients are considered the algorithms' initial population. The GMDH parameters are encoded to start the optimization process.

6- The GMDH model is run. The objective function is computed to assess the performance of GMDH model.

7- The location of agents (particles and honey badgers) shows the values of GMDH parameters. The advanced operators of algorithms are used to change the location of agents. When the locations of agents are updated, the new values of parameters are obtained.

8- The convergence criterion is controlled. If CC is met, the model goes to the testing level; otherwise, it goes to step 6 (Figure 1f).

**2.7 Inclusive multiple model (IMM)**

The advantages and disadvantages of an individual model are numerous. Every model has its advantages and disadvantages. For example, individual models can easily be implemented for predicting variables. They can easily be coupled with optimization algorithms. Studies have shown that individual models are not as accurate as ensemble models (Panahi et al., 2021).

IMM is used to reduce prediction error due to its generalization properties. It is challenging to produce a GMDH model with high precision. The overall precision will increase if modelers integrate multiple GMDH models. This study combined the output of multiple hybrids and standalone GMDH models using an inclusive multiple model. First, outputs were obtained for individual models, including GMDH, GMDH-HBA, GMDH-RSOA, and GMDH-PSOA. The previous level's outputs were then incorporated into a GMDH model. A GMDH model creates synergy between multiple models at this level. GMDH combined the advantages of multiple models to generate the final output.

## 7. Case study

This study aims to determine how coastal vegetation can decrease sediment transport. Therefore, the waves examined in this article are shallow water waves (relative depth less than 0.5). The experiments were conducted in the hydraulic laboratory of Shahrekord University. Flume length, width, and height were 20 m, 60 cm, and 60 cm, respectively. Metal and Plexiglas were used to construct the flume's floor and wall. Figures 2a and 2b show the structure of the flume. The flume was divided longitudinally into sections of 2m, 3m, and 3.6m to construct the water tank, shore, and downstream. The palm tree is resistant tree to the destructive effects of waves. As a result, rigid plastic cylinders were used to simulate vegetation cover on a scale of 1:50. Table 2 shows the details of the experiments. In this study, solitary waves were used to predict STR. The vegetation covers locate based on rectangular and triangular arrangements.

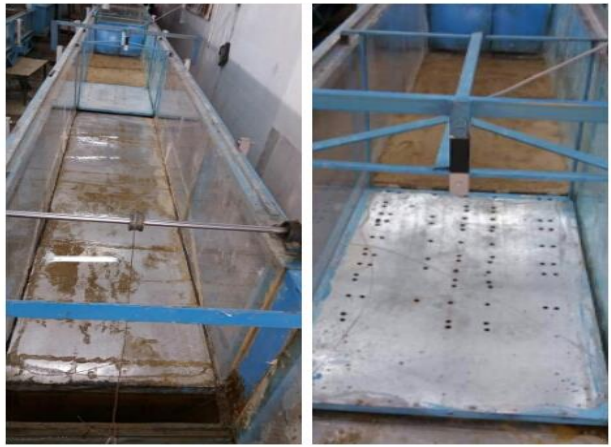

(a)

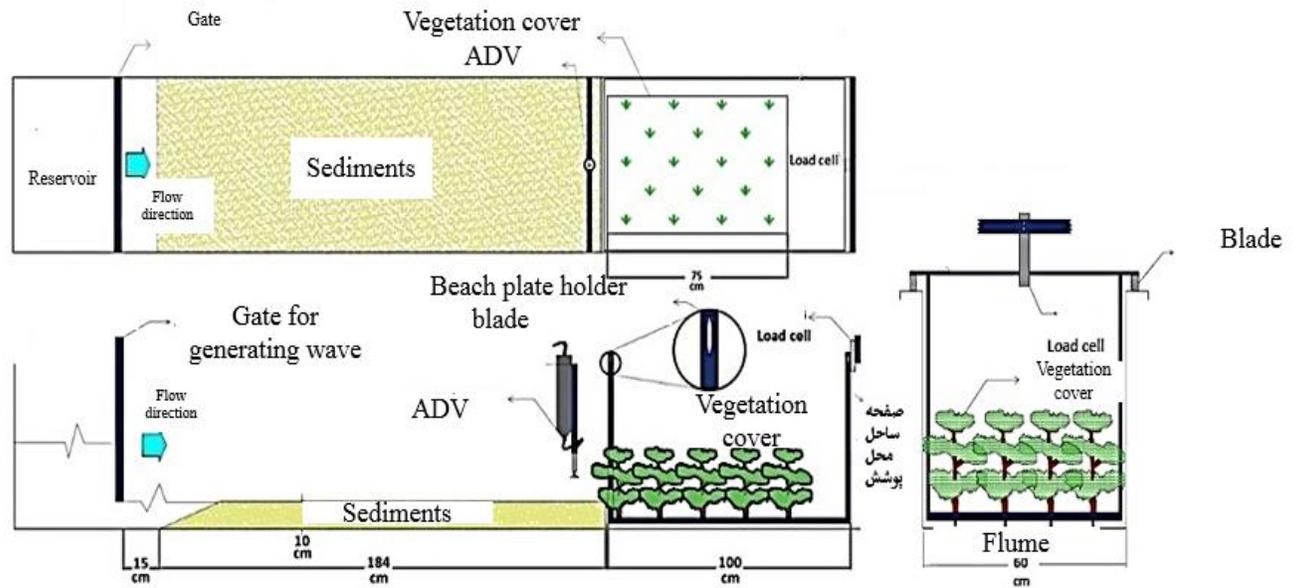

Figure 2. a: Details of used Flume, and b: The structure of flume

Table 2. The details of the layout of the vegetation cover

| Longitudinal and transverse distance | Number of rows | Density (number /m²) | Number of stems | | Configuration (Rectangular) | Configuration (Triangular) |
| --- | --- | --- | --- | --- | --- | --- |
| | | | Rectangularity | Triangular | | |
| 20×20 | 4 | 24 | 12 | 10 | $R_1$ | $T_1$ |
| | 3 | 18 | 9 | 8 | | |
| | 2 | 12 | 6 | 5 | | |
| 15×15 | 5 | 40 | 20 | 18 | $R_2$ | $T_2$ |
| | 4 | 32 | 16 | 14 | | |
| | 3 | 24 | 12 | 11 | | |
| | 2 | 16 | 8 | 7 | | |
| 10×10 | 7 | 77 | 35 | 31 | $R_3$ | $T_3$ |
| | 6 | 66 | 30 | 27 | | |
| | 4 | 44 | 20 | 18 | | |
| | 3 | 33 | 15 | 14 | | |

| | 13 | 273 | 117 | 111 | | |
| --- | --- | --- | --- | --- | --- | --- |
| | 10 | 210 | 90 | 85 | | |
| 5×5 | 7 | 147 | 63 | 60 | $R_4$ | $T_4$ |
| | 4 | 84 | 36 | 34 | | |

The experiment's water distribution system consists of a piping network, a pumping system, and a water tank. The beach was constructed using a galvanized sheet measuring 1 m in length, 0.59 m in width, and 6 mm in thickness. The beach was constructed on a fixed and horizontal slope.

A Plexiglas plate was used to create the tank at the start of the desired area. Afterward, the sliding gate was placed 2 m away from the Plexiglas wall. Based on the height of the tsunami wave, the initial height of the input waves was simulated using a scale of 50:1. Water was pumped into the tank until the desired depth was reached. The gate was quickly opened at the shore, and the broken wave's height was measured. This experiment used two gates. The upstream gate generates waves. The downstream gate balances the water level.

During the refraction moment, the video camera and an ADV (Acoustic Doppler velocimeter) were used to record the wave characteristics. Wave velocity was recorded using the ADV. . The constant sill height (Y) was 6.5 cm.

A dynamometer connected to the transverse part of the flume was applied to record the wave force crashed onto the beach, with and without cover vegetation.

Wave force was recorded with and without cover vegetation using a dynamometer connected to the flume's transverse part.

A preliminary experiment determined that three waves with heights of 25.6 cm, 39.5 cm, and 47 cm behind the upstream gate generate waves with heights of 6, 9, and 12 cm on the beach following the gate. Waves of 6, 9, and 12 cm correspond to refraction moments. During refraction, sediment transport rates depend on wave heights (Jalil-Masir et al., 2021). Table 2 and Figure 3a show details of the forest cover. Sediment transport was examined using two triangular and rectangular forest cover layouts. Figures 3b and 3c show the rectangular and triangular layouts for the vegetation cover density (VCD) =273 (number/m2) and 66 (number/m2). Figure 3d shows sediment transport rates. Erodible materials should cover the flume floor to a certain depth, considered 10 cm. Erodible materials should cover the flume floor to a certain depth, considered 10 cm. As most of the sediments that form beaches are made of sand, sand particles with an average diameter of 0.35 mm are used as porous sedimentary materials in this study. Although drain valves were present, sediments were collected using a net since the model was 3.5m far from the canal's end. Finally, the sediments collected from the drain valve are added to these sediments. A sensitive scale with an accuracy of 0.001 was used to measure the weight of samples after drying samples in a laboratory. For estimating sediment transfer rates, the same conditions were used in different experiments, and the sediments were collected 120 seconds after the experiment began. Based on the unit conversion and the time of the experiment (120s), the rate of sediment transfer was estimated.

Furthermore, Figure 3e shows the heights produced at the refraction moment for VCD=273. This study used sedimentary porous materials with an average diameter of 0.35 mm. Sediment transport rates were also predicted using IMMs, hybrids, and standalone GMDH models.

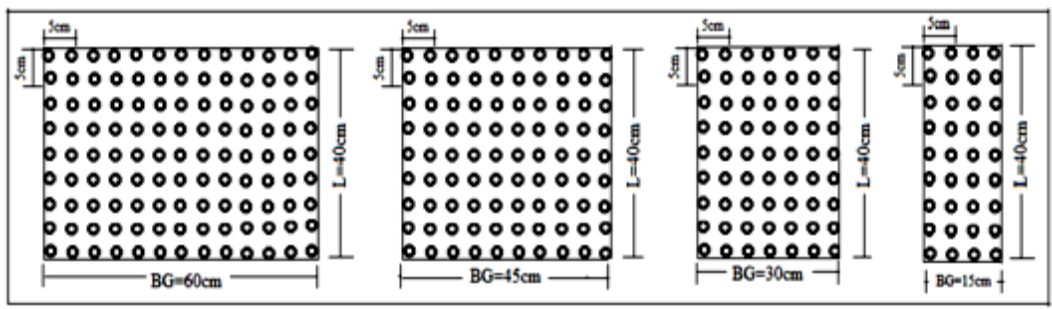

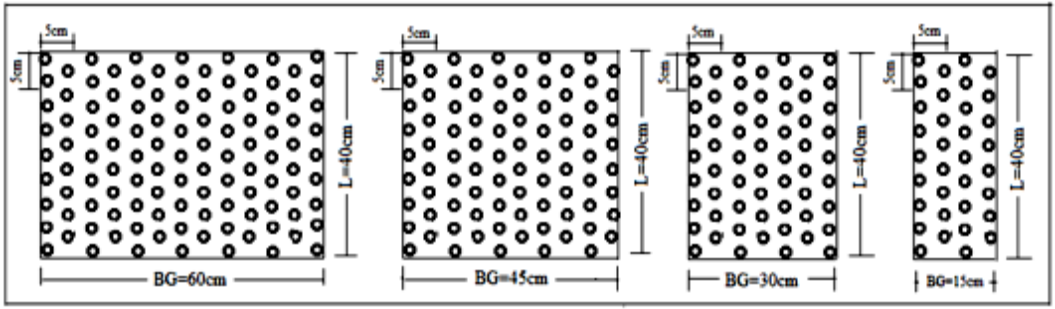

a

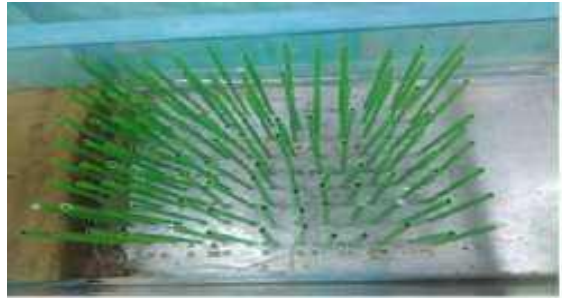

b

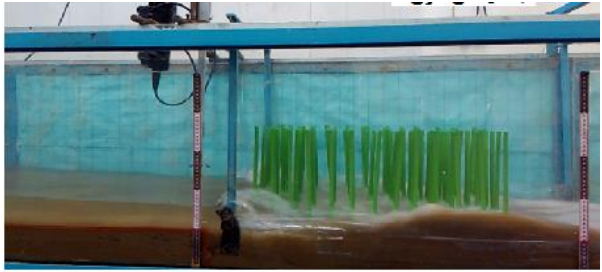

c

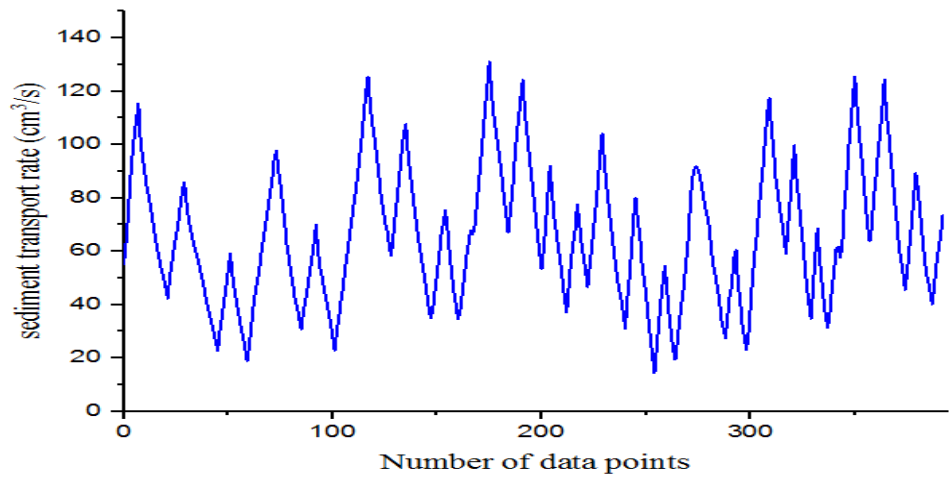

d

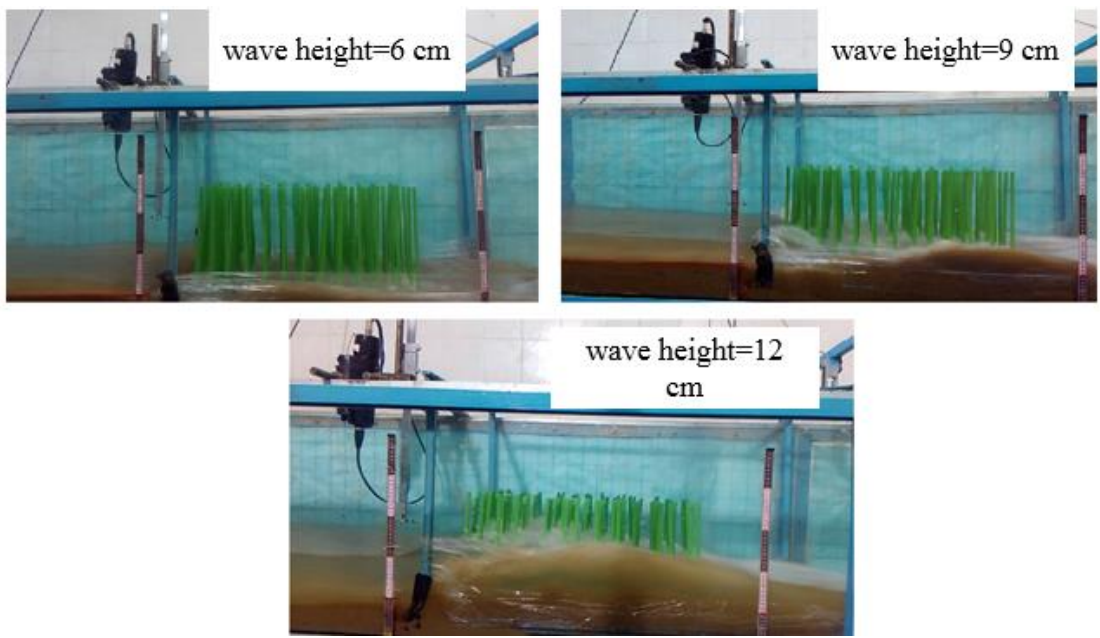

e

Figure 3. a: Configurations of forest covers, b: rectangular layouts for the vegetation cover density (VCD) =273, c: Triangular and triangular arrangements for the vegetation cover density (VCD) =273 and d: sediment transport data points, e: produced waves at the refraction moment

According to Jalil-Masir et al. (2021 and 2022), the sediment transport rate was determined by a variety of factors, including the diameter of the sediments, the diameter of the stems, the cover density, the initial height of the wave, the wave velocity, the cover height, and the wave force. The input parameters in Table 3 were used to predict the sediment transport rate under the specified vegetation conditions. The sediment transport rate was predicted in this study using 393 data sets.

In this study, the following levels were considered for the experiment:

1- In this study, the sand particles were used as the sediment material. The sand was used as the bed material for the flume.
2- Sediment was collected through nets placed after the sediment drain gate.
3- The upstream reservoir was filled with water.
4- The camera and velocity meter were located on the beach for recording information.
5- A pulley and weight system were applied to open the upstream gate abruptly. As a result, a dynamic load sensor (load cell) has been used to directly measure the force applied to the vegetation cover located on the horizontal coast. An electronic display device is connected to the force sensor via connecting wires. Using the load cell, the electronic display senses weight changes based on the voltage change caused by the incoming load.
6- The solitary wave was generated based on an abrupt opening gate.
7- The camera and velocity meter recorded the wave height and velocity in the refraction point. Audio Doppler Velocimeters (ADVs) can be used to check turbulence characteristics and to determine flow patterns and velocities.
8- The sediment samples were gathered downstream.
9- The samples were stored in the laboratory at 25 °C for 36 h.

10- A sensitive balance was used to determine the weights of samples.

The previous studies were investigated to choose the effective input parameters. Based on the previous papers (Chen et al., 2018; Fathi-Moghadam et al., 2018; Jalil-Masir), the parameters of table 3 were chosen for predicting STR. The instantaneous velocity is measured and reported in the table 3.

Table 3. The used inputs for predicting sediment transport rate

| Parameter | Average | Maximum | Minimum | Standard deviation |
|---|---|---|---|---|
| Number of experiments: 393 and number of data:393 | | | | |
| height at the refraction moment ($H_W$) (cm) (height wave after gate) | 9 | 12 | 6 | 2.64 |
| Vegetation cover density (number /m²) (DS) | 73.33 | 273 | 12 | 75.44 |
| Wave force (F) (unit:N) | 51.66 | 190.29 | 12.86 | 33.50 |
| $D_{50}$ (mm) | 0.33 | 0.35 | 0.30 | 0.26 |
| Height of vegetation cover ($h_v$) (cm) | 32 | 35 | 30 | 5.73 |
| cover stem diameter D (cm) | 0.55 | 0.90 | 0.30 | 0.40 |
| Velocity (m/s) $V_W$ | 1.46 | 1.53 | 1.34 | 0.03 |

The following indices were used as follows (Bazrafshan et al., 2022):

1- Root mean square error

$$RMSE = \sqrt{\frac{\sum_{i=1}^{N}(RST_{ob} - RST_{es})^2}{N}} \qquad (22)$$

2- Mean absolute error (MAE)

$$MAE = \frac{1}{N}\sum_{i=1}^{N}|RST_{ob} - RST_{es}| \qquad (23)$$

3- Percentage of bias (PBIAS)

$$PBIAS = \frac{\sum_{i=1}^{N}(RST_{es} - RST_{ob})}{\sum_{i=1}^{N}RST_{ob}} \qquad (24)$$

4- Nash–Sutcliffe efficiency

$$NSE = 1 - \frac{\sum_{i=1}^{N}(RST_{es} - RST_{ob})^2}{\sum_{i=1}^{N}(RST_{ob} - \overline{RST}_{ob})^2} \qquad (25)$$

5- Fraction of standard deviation (FSD):

$$FSD(RST_{es}, RST_{ob}) = 2 * \frac{|SD(RST_{es}) - SD(RST_{ob})|}{SD(RST_{es}) + SD(RST_{ob})} \qquad (26)$$

Where RST: rate of sediment transport, $SD(RST_{es})$ SD: standard deviation of estimated RST, $SD(RST_{ob})$: standard deviation of observed RST, $RST_{obs}$: observed RST, $RST_{es}$: estimated RST, $\overline{RST}_{ob}$: Average observed RST, N: number of data.

## 8. Results and Discussion

### 4.1 Training and testing level size selection

Figure 4 illustrates the various data sizes tested for the individual models at the training and testing levels to determine the optimal size. The optimal training and testing data sizes for all models were 70% and 30%, respectively, since RMSE had the smallest value for these data sizes. Based on the GMDH model, the objective function for 50%, 55%, 60%, 65%, 70%, and 80% of data was 1.05, 0.91, 0.967, 0.845, 0.800, and 0.833, respectively. This study selected 70% and 30% of the data for training and testing levels, respectively.

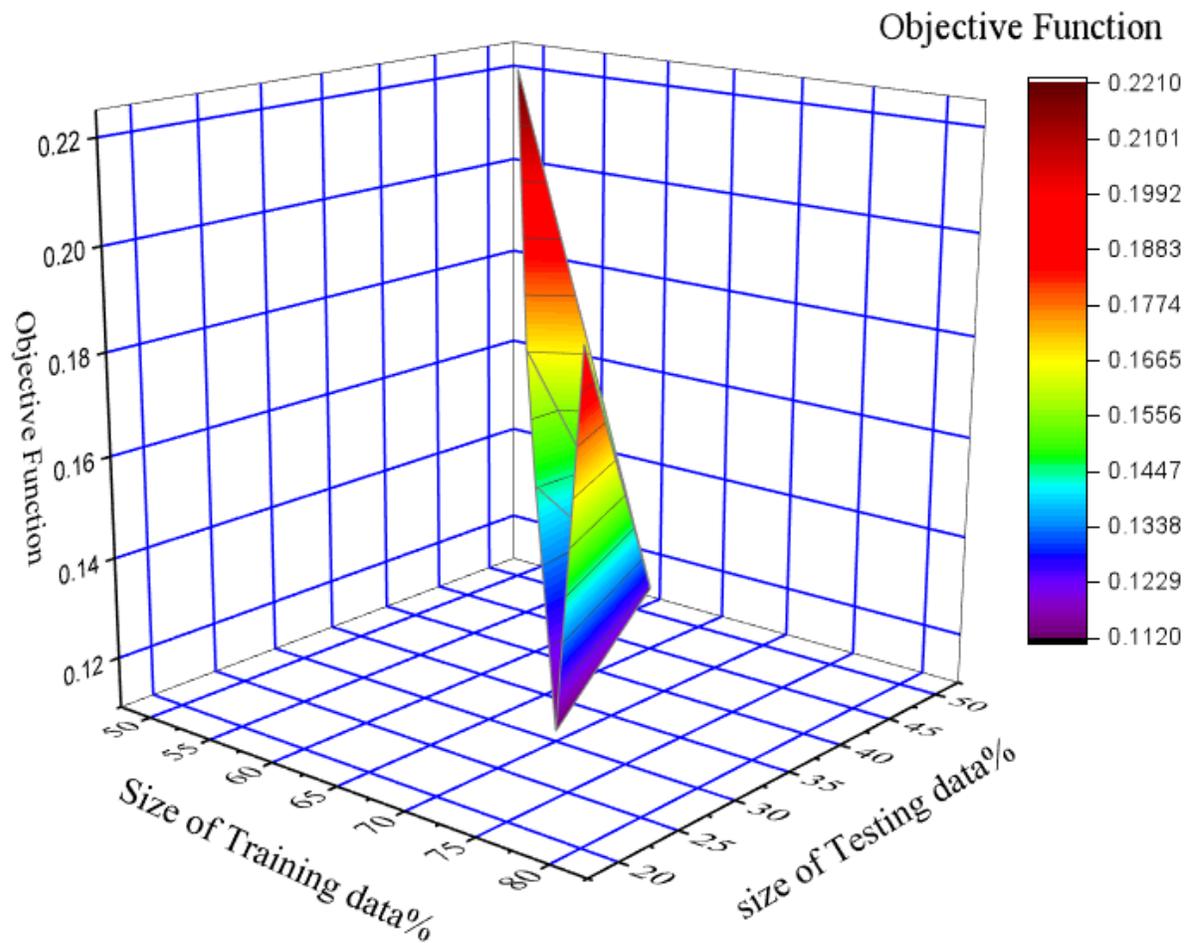

HBA

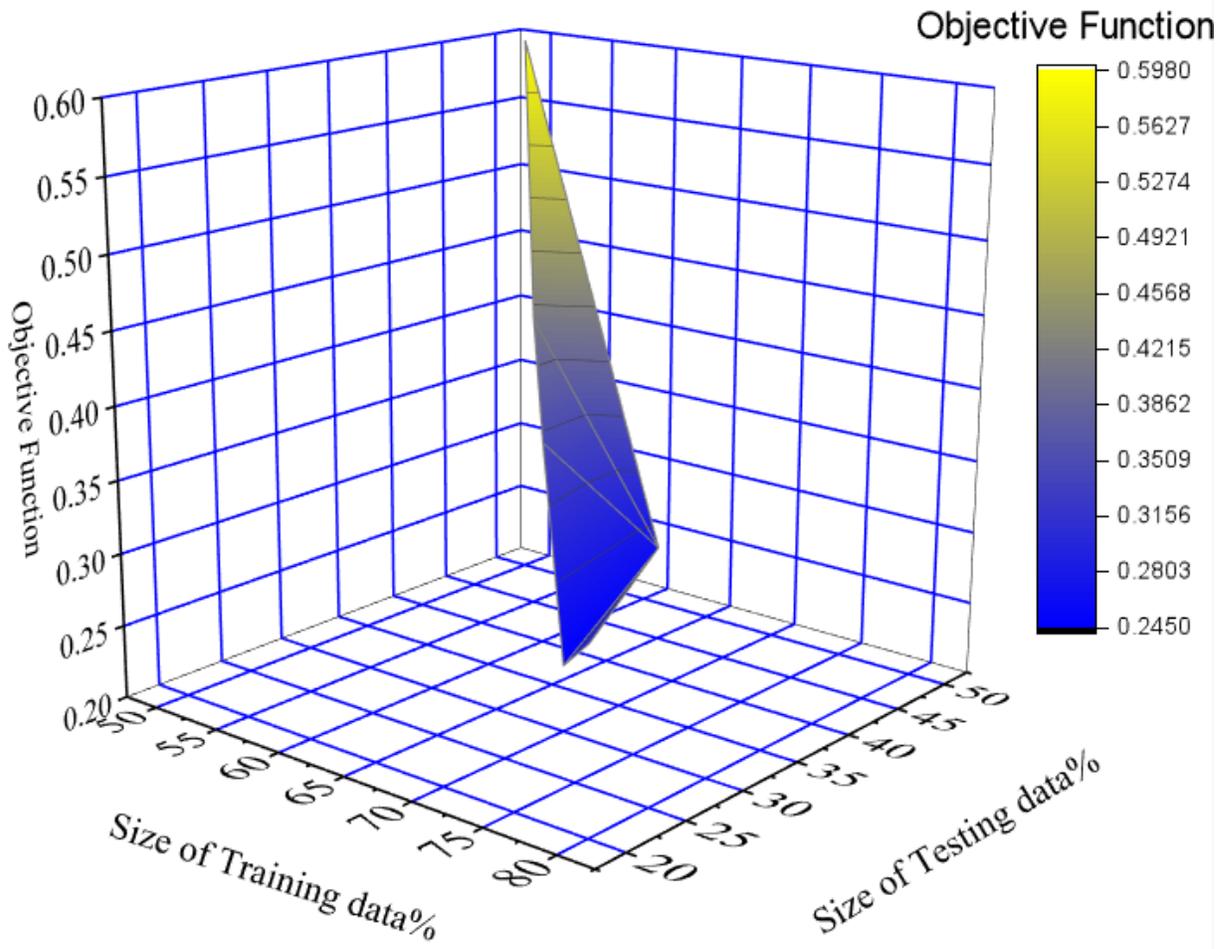

RSOA

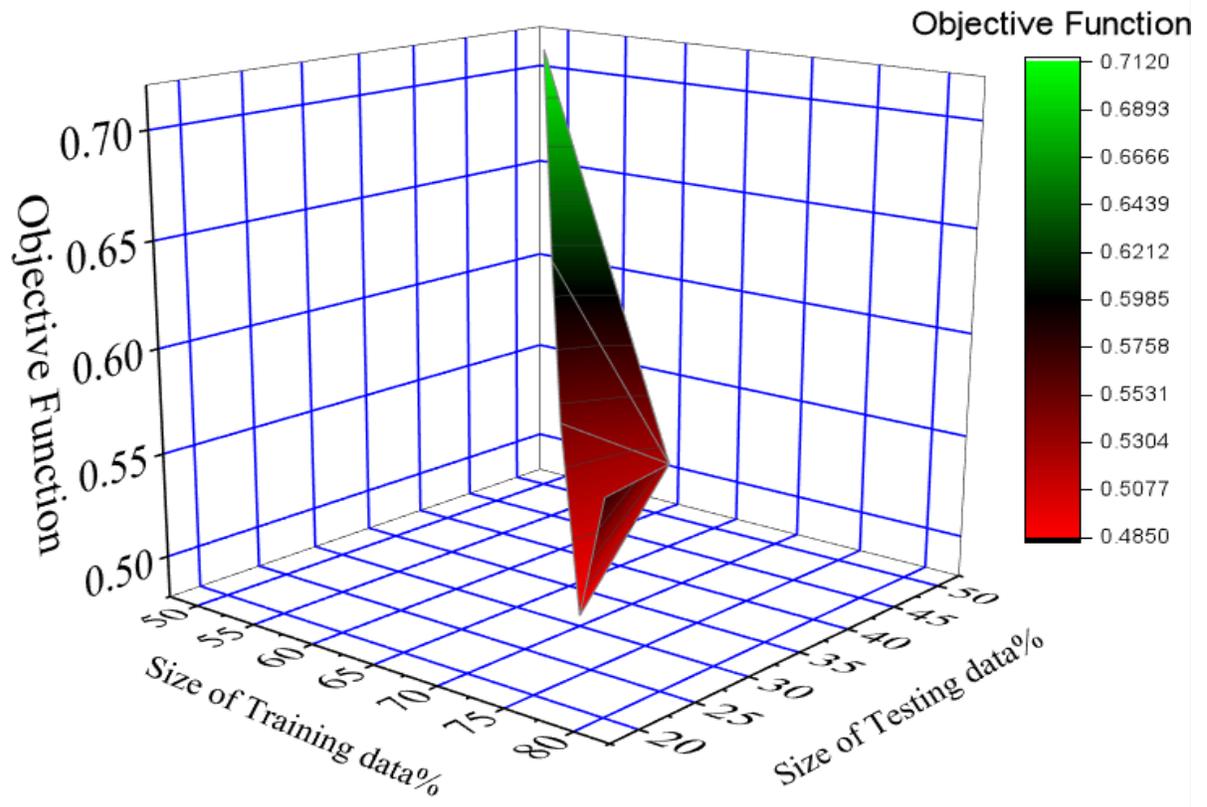

SCA

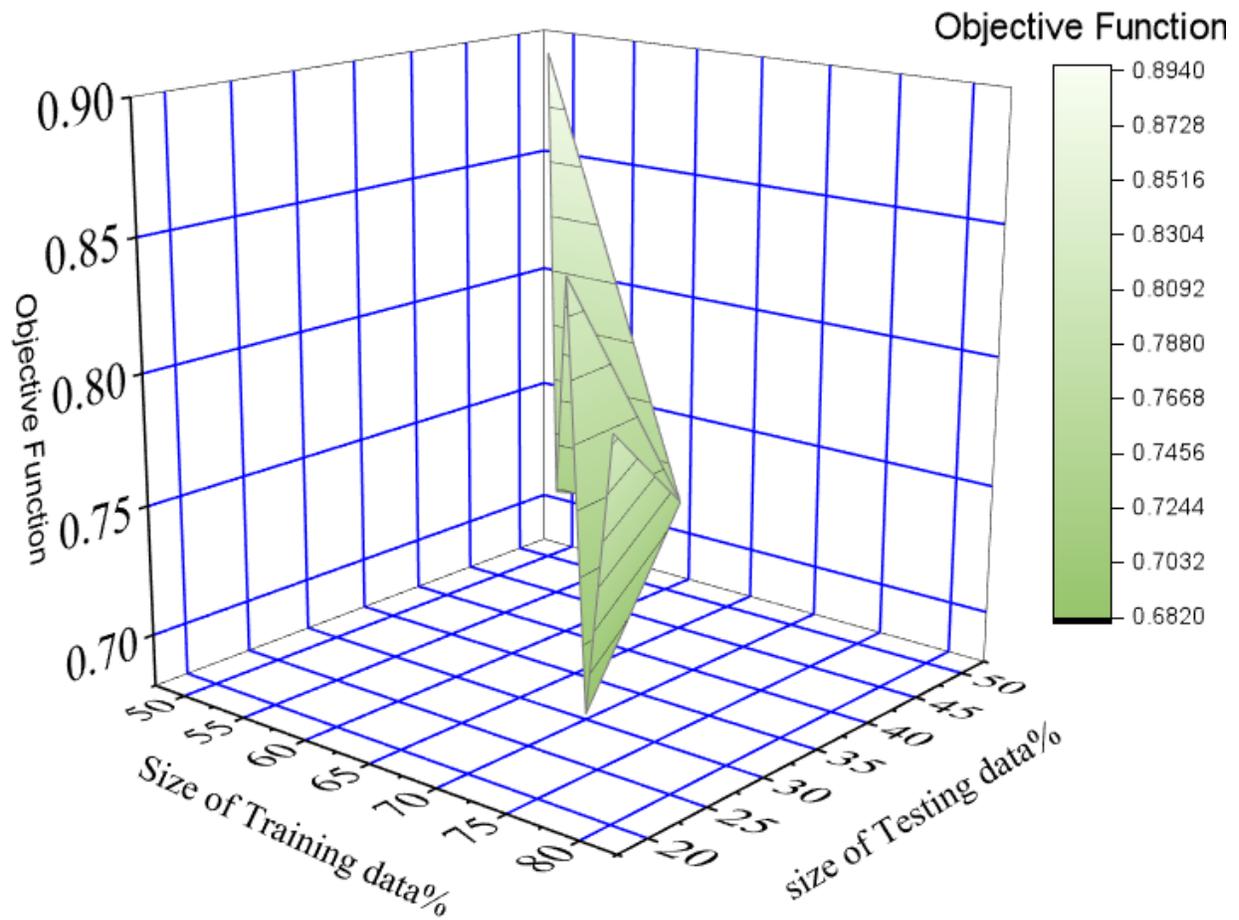

PSOA

Figure 4. The choice of best size for data

**4.2 Best value determination for random parameters**

Optimization algorithms have random parameters (RPs). It is important to determine RPs optimally when using optimization algorithms. RP affects the objective function, as shown in Table 3. The HBA population size varied from 50 to 200. The population size of 100 produced the smallest objective function value (RMSE). HBA iterations ranged from 40 to 160. Based on the objective function value (RMSE), 80 was the best maximum number of iterations for HBA. The population size of RSOA varied from 50 to 200. 100 was the best population size for RSOA. The RSOA achieved the lowest objective function value at 100 iterations. Similar processes were used to determine the values of other algorithm parameters. Other parameters were left unchanged, while one parameter was varied.

Table 4. The sensitivity analysis of random parameters of algorithms

(Population size: POS, Objective Function: OBFU, the maximum number of iterations: MANI)

HBA

| POS | OBFU | MANI | OBFU |
|---|---|---|---|
| 50 | 0.167 | 40 | 0.169 |
| 100 | 0.123 | 80 | 0.134 |
| 150 | 0.190 | 120 | 0.187 |
| 200 | 0.198 | 160 | 0.98 |

RSOA

| POS | OBFU | MANI | OBFU |
|---|---|---|---|
| 50 | 0.345 | 50 | 0.367 |
| 100 | 0.267 | 100 | 0.266 |
| 150 | 0.298 | 150 | 0.399 |

| | | | | |
|---|---|---|---|---|
| 200 | 0.312 | 200 | 0.412 | |

## SCA

| POS | OBFU | Maximum number of iterations | OBFU | $\kappa_2$ | OBFU | $\kappa_2$ | OBFU |
|---|---|---|---|---|---|---|---|
| 50 | 0.345 | 100 | 0.367 | π/3 | 0.389 | 0.60 | 0.376 |
| 100 | 0.267 | 200 | 0.266 | 2π/3 | 0.276 | 0.80 | 0.265 |
| 150 | 0.298 | 300 | 0.399 | 3π/3 | 0.265 | 1.00 | 0.298 |
| 200 | 0.312 | 400 | 0.412 | 4π/3 | 0.291 | 1.2 | 0.322 |

## PSOA

| POS | OBFU | MANI | OBFU | $\mu_1 = \mu_2$ | OBFU | $\phi$ | OBFU |
|---|---|---|---|---|---|---|---|
| 100 | 0.991 | 100 | 0.867 | 1.60 | 0.865 | 0.30 | 0.767 |
| 200 | 0.782 | 200 | 0.712 | 1.80 | 0.787 | 0.50 | 0.687 |
| 300 | 0.682 | 300 | 0.667 | 2.00 | 0.687 | 0.70 | 0.871 |
| 400 | 0.724 | 400 | 0.694 | 2.200 | 0.623 | 0.90 | 0.891 |

**4.3 Best input scenario selection**

In this section, eight input scenarios were considered to determine the effects of inputs on outputs. Table 5 defines the input scenario. The objective function (RMSE) for various input scenarios and models is illustrated in Figure 5. Based on input scenarios (1) -(8), the RMSE of GMDH-HBA was 0.123 cm3/s, 0.233 cm3/s, 0.322 cm3/s, 0.412 cm3/s, 0.523 cm3/s, 0.612 cm3/s, 0.789 cm3/s, and 0.812 cm3/s. Using all input variables produced the best results. Eliminating the wave height at the refraction moment from the input combination increased RMSE by 84%. Removing the cover height from the input combination increased RMSE by 47%. The wave height at the refraction moment and cover height showed the highest and lowest significance for the GMDH-HBA, respectively.

Based on the input scenarios (1) -(8), GMDH-RSOA had RMSEs of 0.267 cm3/s, 0.278 cm3/s, 0.392 cm3/s, 0.567 cm3/s, 0.612 cm3/s, 0.823 cm3/s, 0.901 cm3/s, 0.911 cm3/s. Removing wave height from the input combination increased RMSE by 70%. For GMDH-RSOA, wave velocity was another important parameter. Various models indicated that density cover, wave height, and D50 were the most important parameters for predicting sediment transport rate.

Table 5. The defined input scenarios for the models

| Input scenario | |
|---|---|
| 1 | H (W), V(W), D(S), F, $D_{50}$, h(v), D |
| 2 | H (W), V(W), D(S), F, $D_{(50)}$, D |
| 3 | H (W), V(W), D(S), F, $D_{(50)}$, h(v) |
| 4 | H (W), V(W), D(S), $D_{(50)}$, h(v) , D |
| 5 | H (W), V(W), D(S), F, h(v), D |
| 6 | H (W), V(W), F, $D_{(50)}$, h(v), D |
| 7 | H (W), D(S), F, $D_{(50)}$, h(v), D |

 V(W), D(S), F, $D_{(50)}$, h(v), D

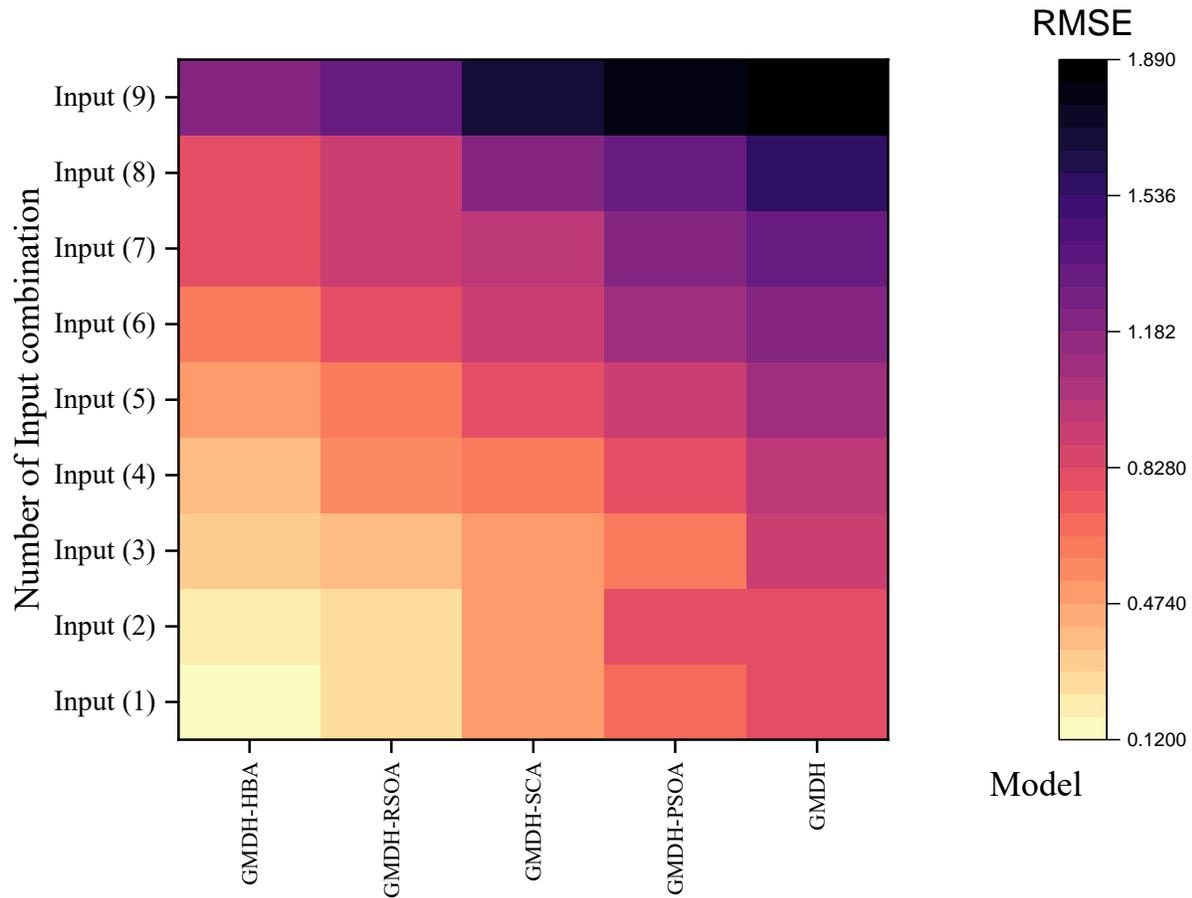

Figure 5. The computed RMSE for different input scenarios and models

## 4.4 Model accuracy evaluation

The first input scenario was used to run all models in this section. Figure 6 illustrates radar plots to assess the accuracy of models. The accuracy of models based on PBIAS is evaluated in Figure 6a. The IMM achieved a PBIAS of 8%, whereas PBIASs of 12, 16, 22, 27, and 33% were achieved by the GMDH-HBA, GMDH-RSOA, GMDH-SCA, GMDH-PSO, and GMDH at the training level.

Based on PBIAS, the IMM and GMDH produced the best and worst results, respectively. At the testing level, GMDH-HBA had a PBIAS of 15%. The HBA performed better than the other optimization algorithms. Figure 6b shows the NSE values for the models. At the training level, the IMM had an NSE of 0.98, while the GMDH-HBA, GMDH-RSOA, GMDH-SCA, GMDH-PSO, and GMDH had NSEs of 0.97, 0.94, 0.87, 0.85, and 0.80.

The IMM, GMDH-HBA, and GMDH-RSOA showed the highest accuracy at the testing level. The FSD values for the various models are shown in figure 6c. IMM had a FSD of 1.11, while GMDH-HBA, GMDH-RSOA, GMDH-SCA, GMDH-PSO, and GMDH had FSDs of 0.23, 0.67, 0.85, 1, and 1.22, respectively. GMDH-HBA performed better than GMDH-RSOA, GMDH-PSOA, GMDH-SCA, and GMDH based on the accuracy of the results. The MAE values for the various models are shown in figure 6d. At the training level, the IMM decreased the MAE of the GMDH-HBA, GMDH-RSOA, GMDH-SCA, GMDH-PSO, and GMDH by 24%, 58%, 58%, 75%, and 80%, respectively. At the testing level, the IMM had the lowest MAE. The IMM and GMDH-HBA had the highest accuracy in this section. The HBA uses advanced operators. Equations 10 and 13 can be used to update HBA solutions. In this study, the IMM model performed better than other models. For predicting STR, Jalil-Masir et al. (2021) used regression models and the same data points. They reported the $R^2$ value of 0.84 for the regression model. Thus, the IMM of the current study performed better than the regression.

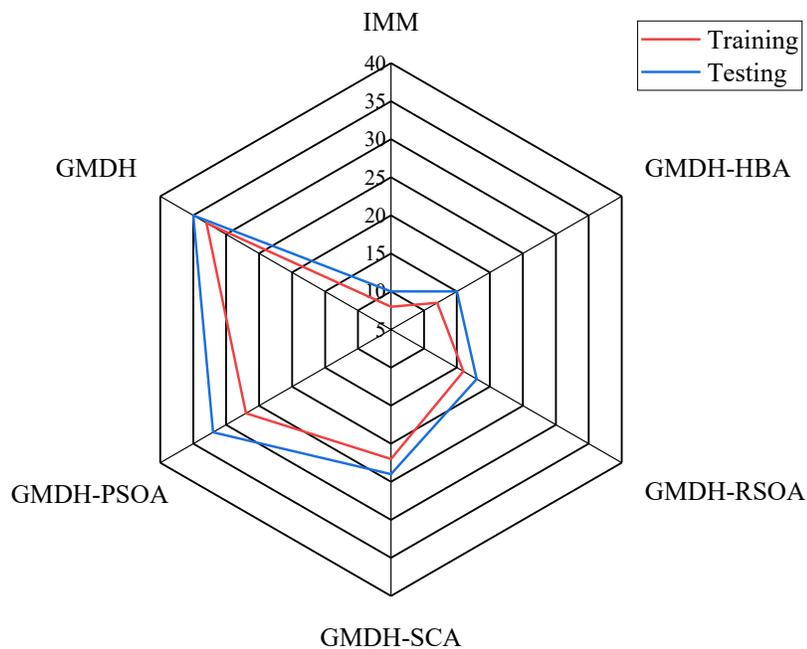

(a)

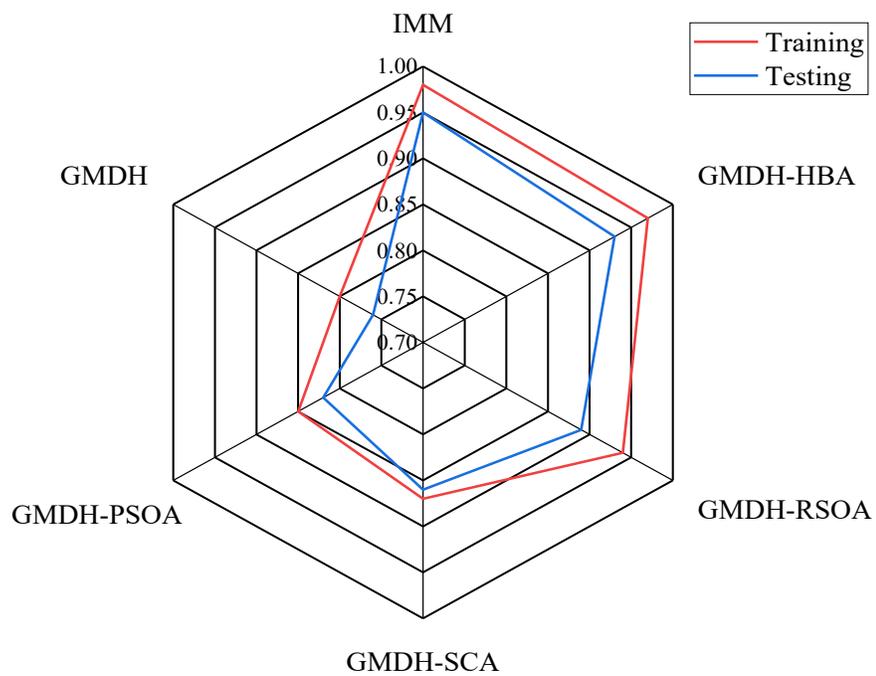

b

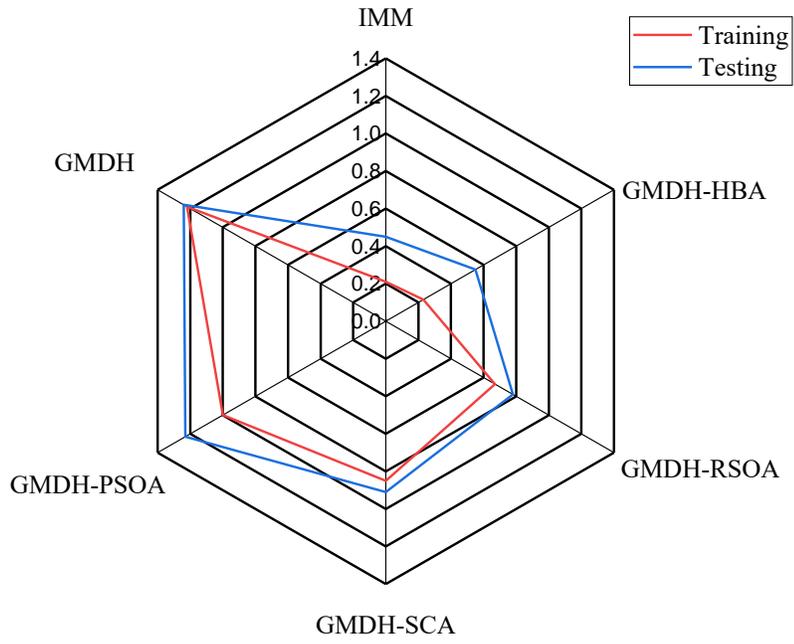

c

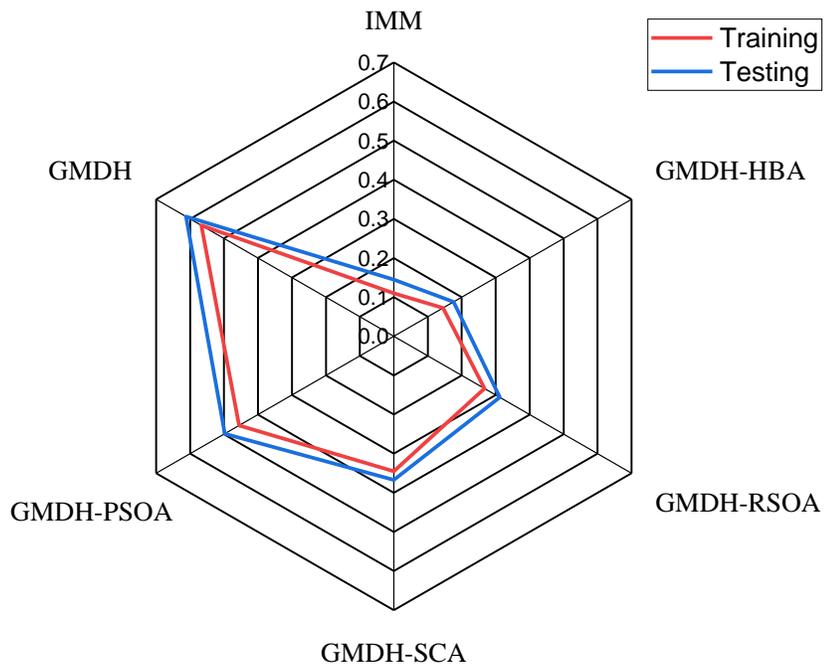

d

Figure 6. Radar plots base for a: PBIAS, b: NSE, c: FSD, and d: MAE

The HBA outperformed the other algorithms for the reasons stated previously. The GMDH was trained using backpropagation, but optimization algorithms significantly improved its accuracy. Figure 7 shows the histogram of the residuals. In Figure 7a, you can see the residual histogram for the IMM model. In this Figure, 308 and 73 data points fall within the -1 and 1 bin centers, respectively. Other bin centers receive a small number of data. Figure 7b shows the residual histogram for the GMDH-HBA model. The residual of 263 data points falls in the 2.5 bin center. Additionally, 112 and 7 data points fall in the -2.5 and -7.5 bin centers, respectively. Figure 7c shows the residual histogram for the GMDH-RSOA model.

The residuals for 301, 60, 12, 4, and 3 data points are 2.5, -2.5, -7.5, -12.5, and -17.5, respectively. Figure 7d shows the residual histogram for the GMDH-SCA model. For 300, 53, 15, 7, and 5 data points, the residuals are 2.5, -2.5, -7.5, -12.5, and -17.5, respectively. Figure 7e shows a histogram of the residual for GMDH-PSOA. Two hundred seventy-two data points, 73 data points, 15 data points, 9 data points, and 5 data points yielded residuals of 2.5, -2.5, -7.5, -12.5, and 17.5, respectively. The residuals of other data points fall into other center bins. A histogram of the residuals of GMDH can be seen in Figure 7f. 309, 25, 17, 12, 8, 5, and 2 data points have residuals of 2.5, -2.5, -7.5, -12.5, -17.5, -22.5, and -27.5, respectively. The residuals of other data points fall into the other center bins. Some of the outputs of GMDH had residuals of >-42.5. The IMM and GMDH outperformed the other models in this section.

Figure 7g shows the boxplots for the models. The median of the observed data, IMM, GMDH-HBA, GMDH-RSOA, GMDH-SCA, GMDH-PSOA, and GMDH were 58.5 cm$^3$/s, 58.5 cm$^3$/s,

58.5 cm³/s, 63 cm³/s, 63 cm³/s, 63 and 63 cm³/s, respectively. The mean of the observed data, IMM, GMDH-HBA, GMDH-RSOA, GMDH-SCA, GMDH-PSOA, and GMDH was 67.0 cm³/s, 67.0 cm³, 66.7 cm³/s, 68.7 cm³/s, 69.0 cm³/s, 69.7 cm³/s, and 69.8 cm³/s, respectively. It was discovered that the IMM and observed data had a high degree of correspondence.

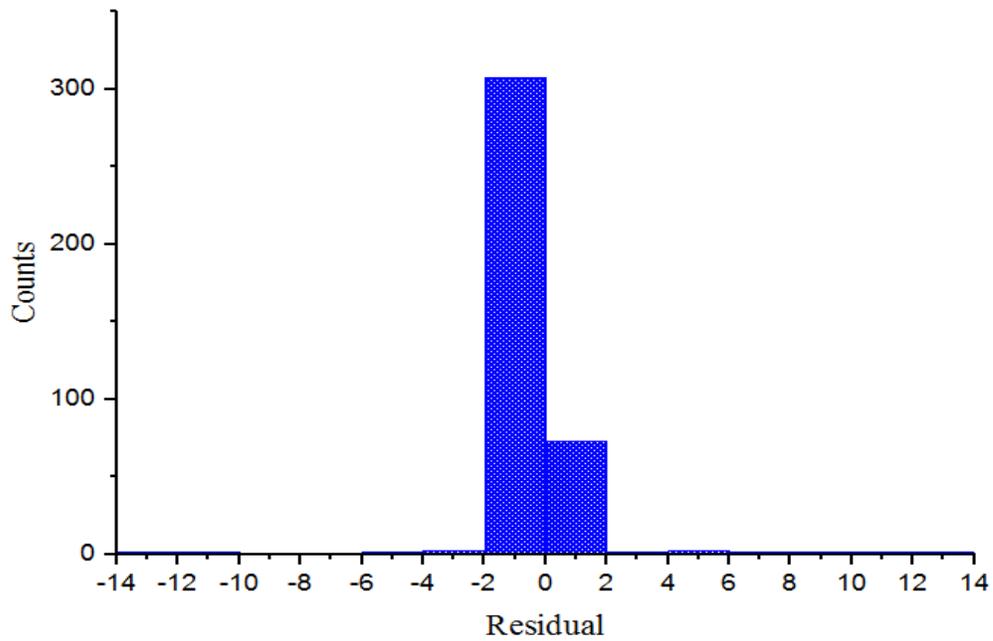

(a)

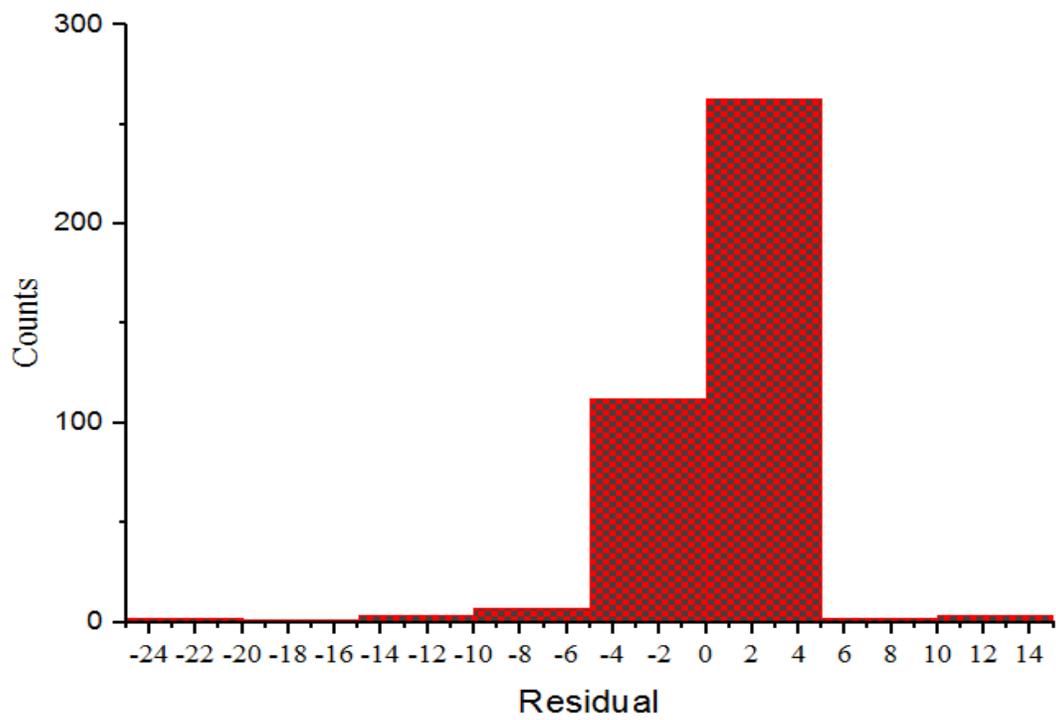

(b)

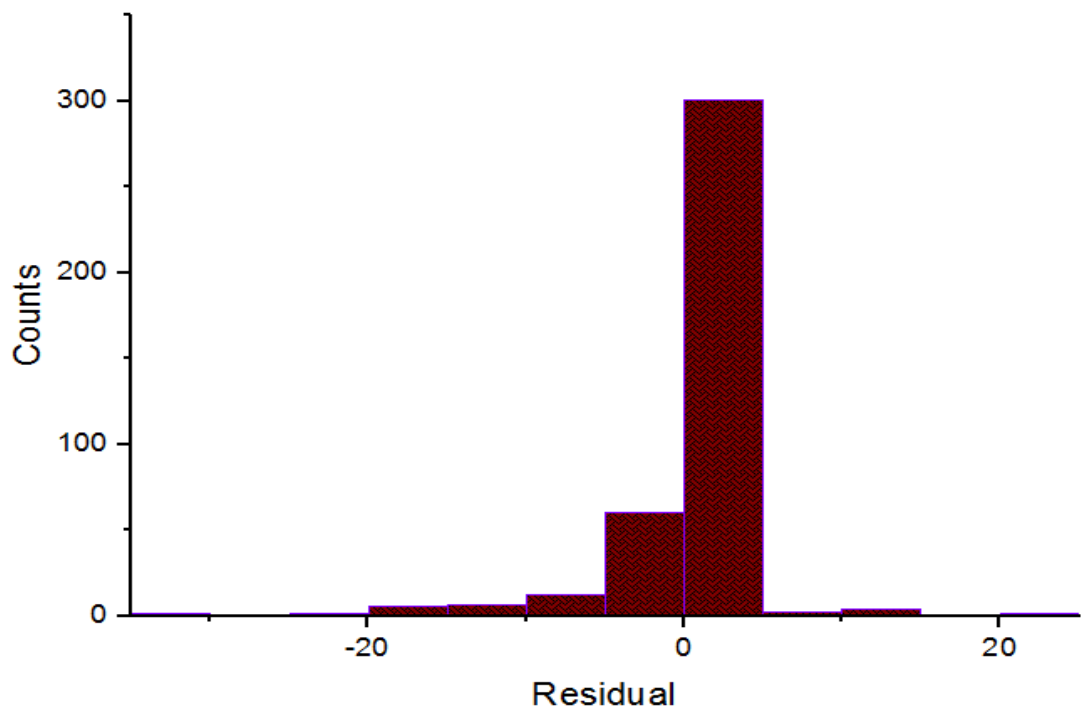

(c)

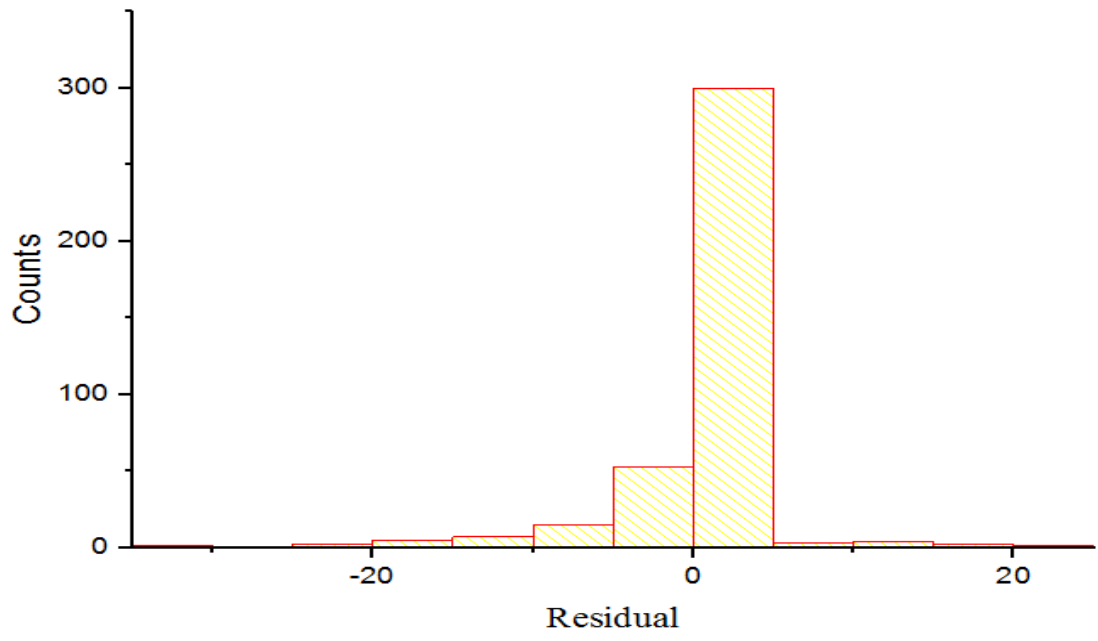

(d)

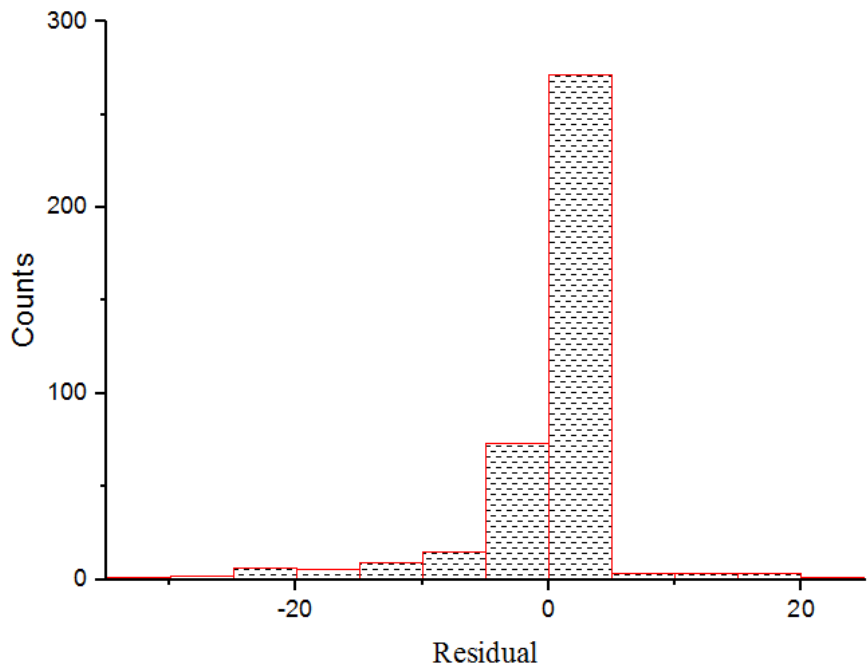

e

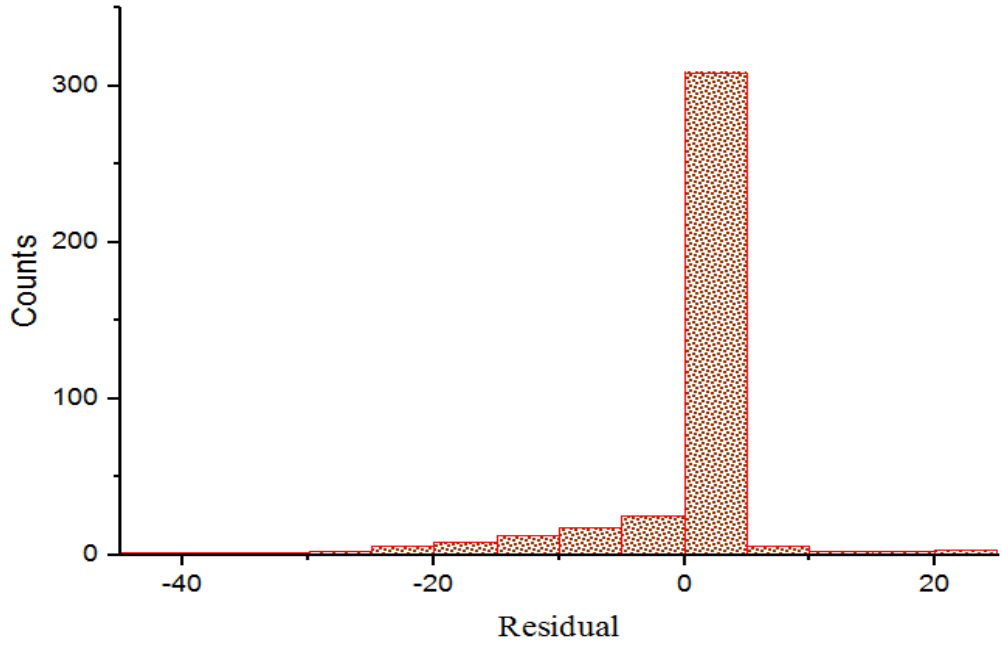

F

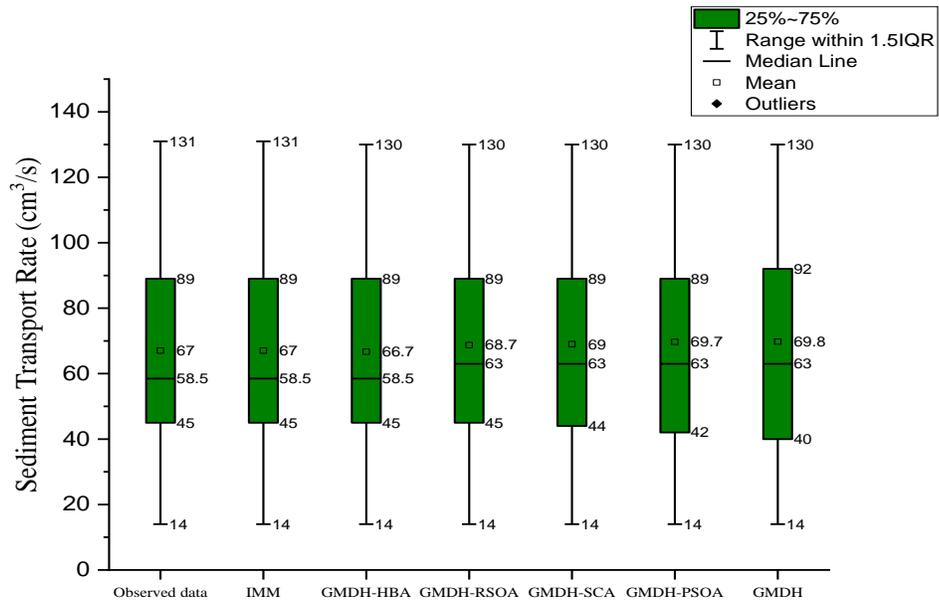

g

Figure 7. The histogram of residual values of models for a: IMM, b: GMDH-HBA, c: GMDH-RSOA, d: GMDH-SCA, e: GMDH-PSOA, f: GMDH, and g: The boxplots of models for estimating STR

Figures 8a, 8b, 8c, 8d, 8e, and 8f show density scatterplots for the IMM, GMDH-HBA, GMDH-RSOA, GMDH-SCA, GMDH-PSOA, and GMDH. The $R^2$ values for IMM, GMDH-HBA, GMDH-RSOA, GMDH-SCA, GMDH-PSOA, and GMDH were 0.9963, 0.9887, 0.9741, 0.9669, 0.9443, and 0.9267, respectively. Based on the density values, the highest density was between 43.80 cm3 and 73.70 cm3, implying that this interval contains the most data. A large number of data points overlap in this region. There is a low-density value at all models' upper and lower limits.

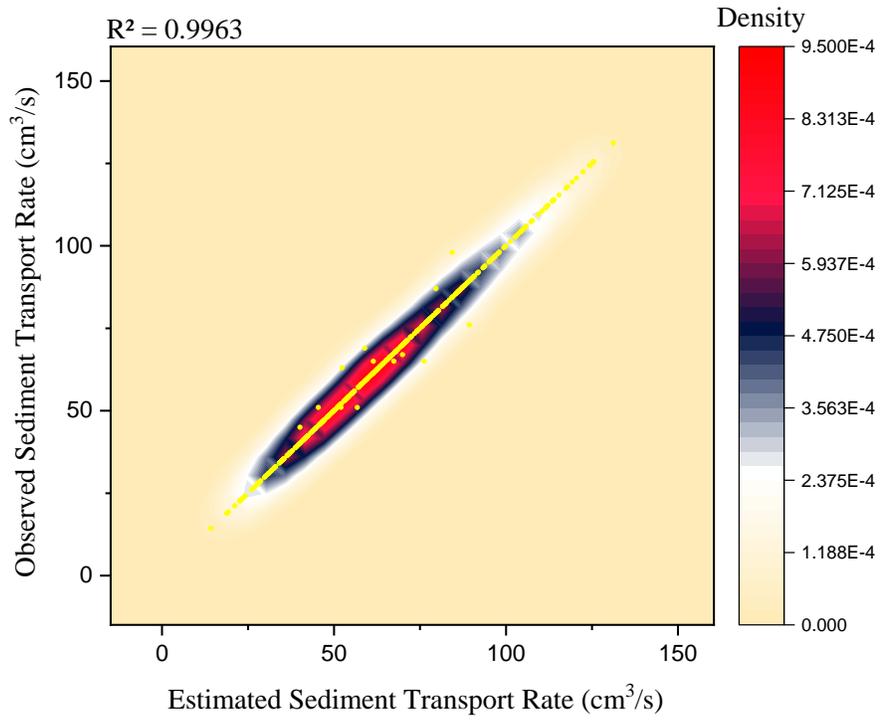

IMM

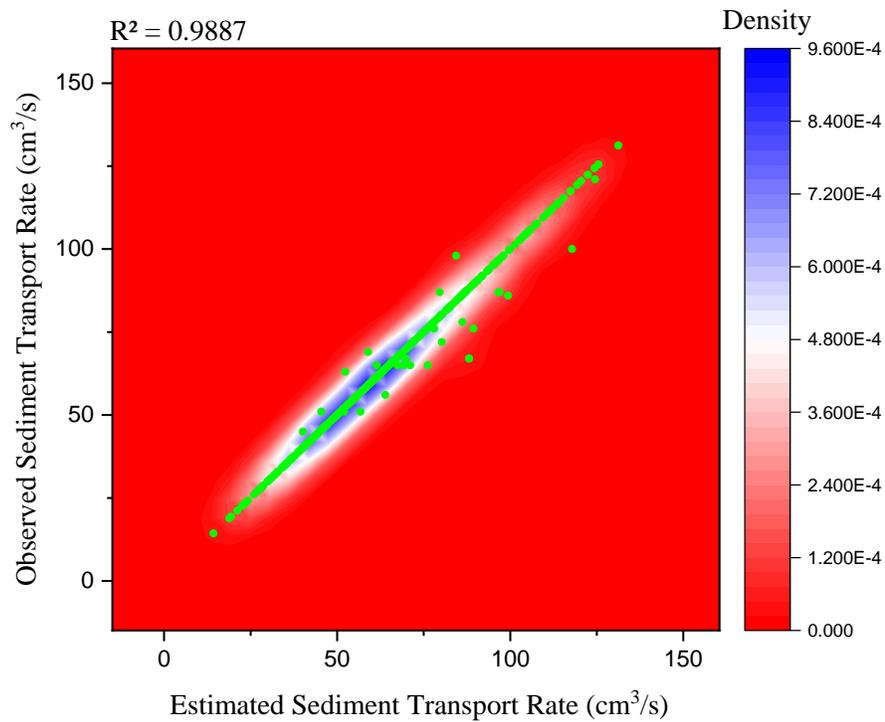

GMDH-HBA

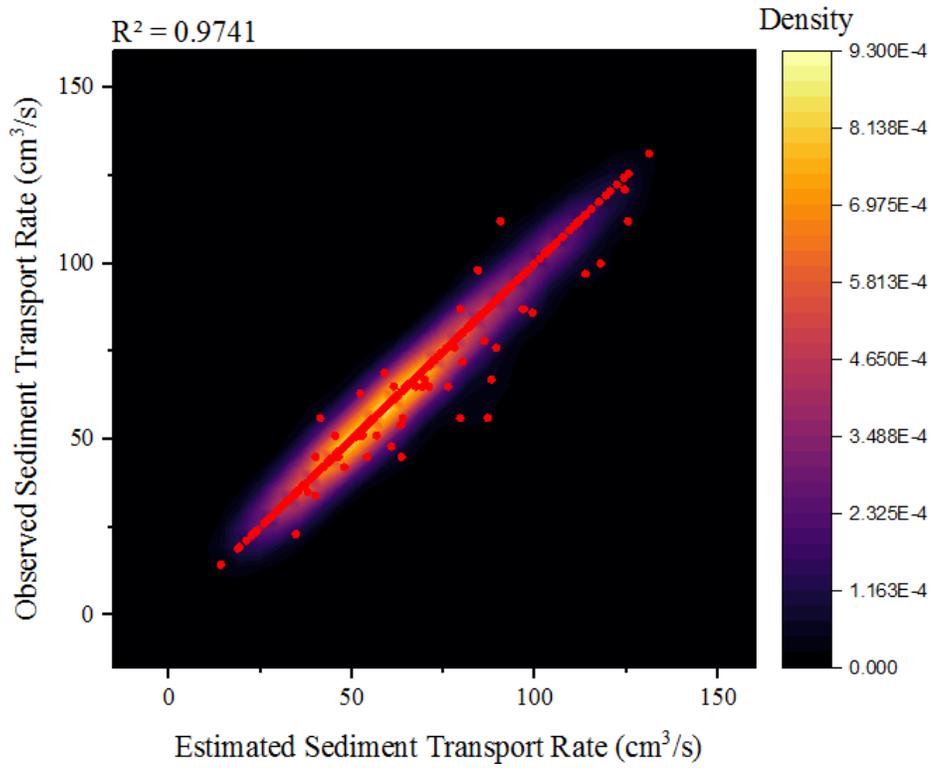

GMDH-RSOA

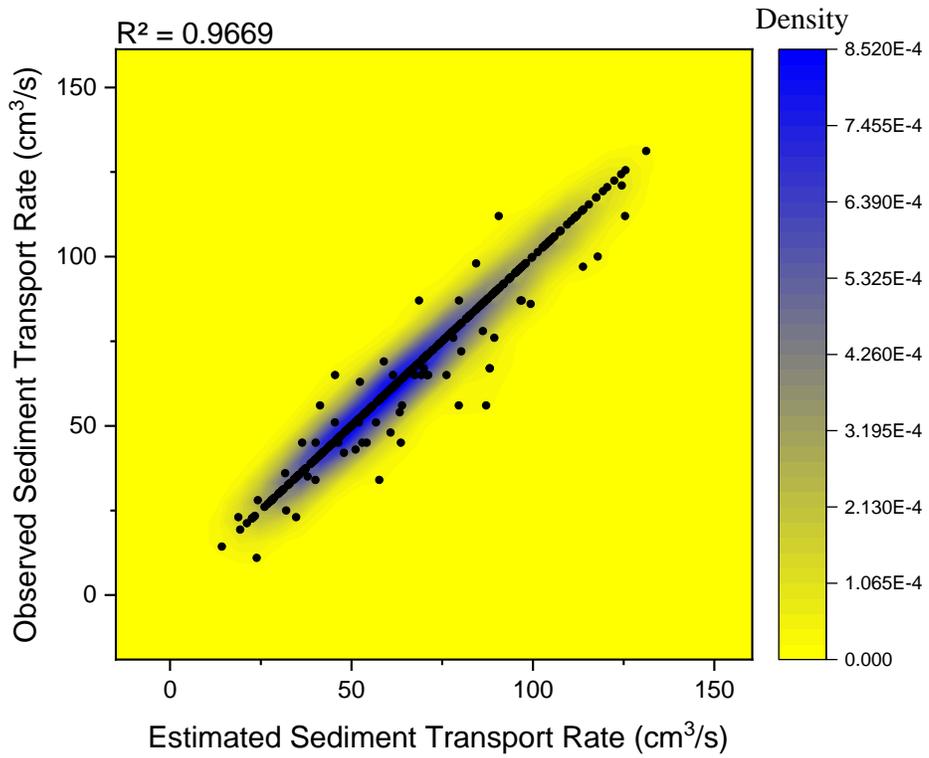

GMDH-SCA

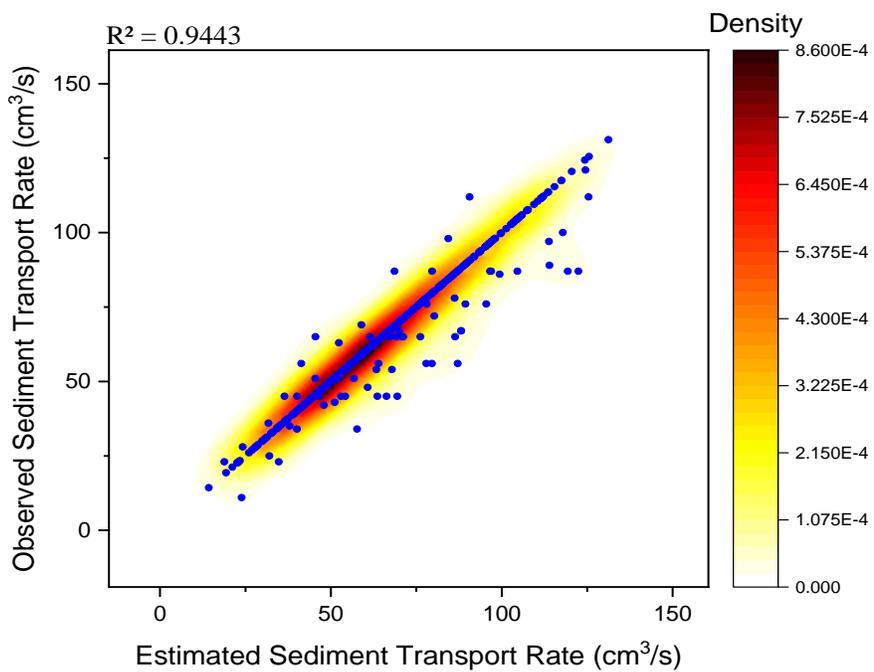

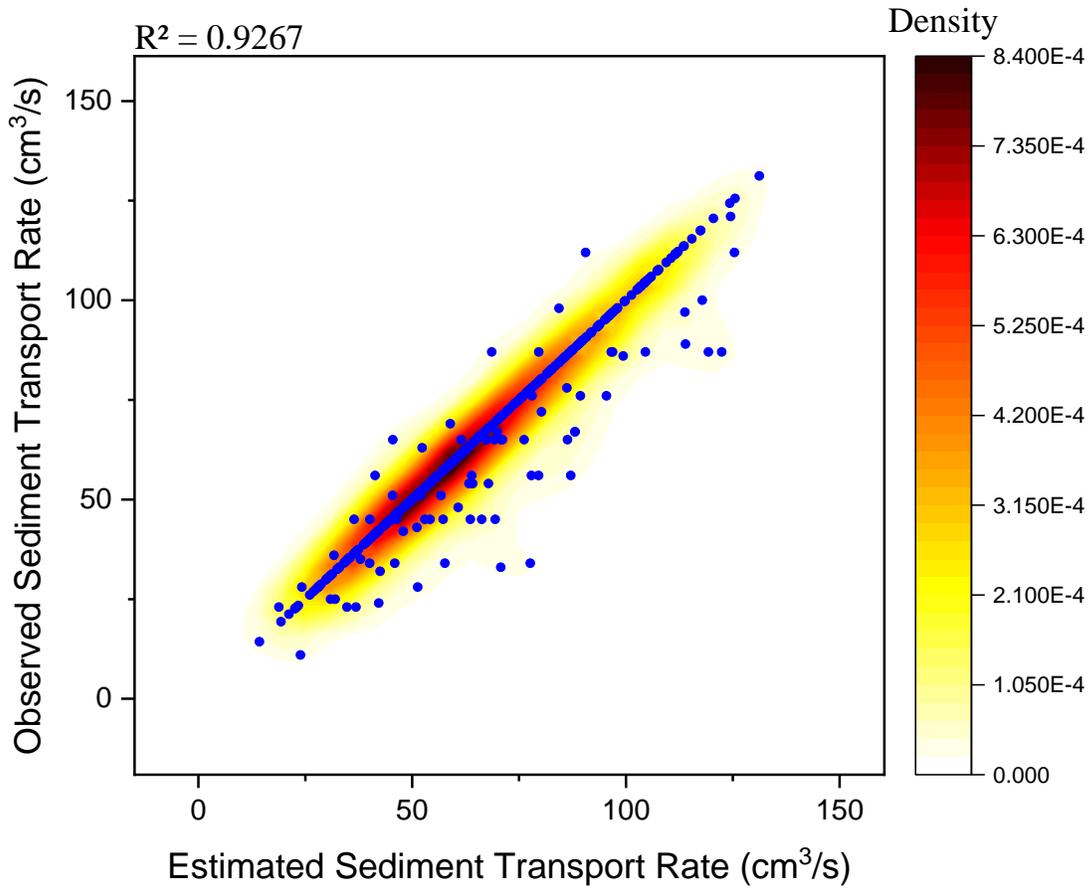

Figure 8. The heat scatter plots of models

## 4.5 Further discussion

The effect of various parameters on sediment transport rate was examined in this section. In this study, experiments were conducted using vegetation cover. Sediment transport and wave velocity are reduced by vegetation cover. Using a rectangular layout, Figure 9a illustrates the velocity and height wave variations for different vegetation cover densities (VCDs). The wave velocity should increase for a constant height wave when the VCD value increases. As a result, forest cover effectively reduces the height wave at the refraction moment. At VCD =12 and wave height =6 cm, the rectangular layout produces a wave velocity of 1.34 m/s. At VCD =44 and VCD =210, the wave velocity should be increased from 1.34 (VCD =12) to 1.40 (VCD =44) m/s and from 1.34 (VCD =12) to 1.42 (VCD =210) m/s for maintaining a constant height of 6 cm. Figure 10b shows the VCD variations for three triangular VCDs of 12, 44, and 210.

As the VCD increased, the wave height decreased in the triangular layout. Figures 9a and 9b demonstrated that the triangular layout reduced height waves more than the rectangular layout. At VCD = 12 and wave height = 6 cm, the rectangular layout produces a wave velocity of 1.34 m/s, while the triangular layout produces a wave velocity of 1.38 m/s. It is necessary to increase the velocity from 1.34 m/s to 1.38 m/s to maintain the 6 cm height. Wave force can be reduced more effectively by a triangular configuration. Thus, wave velocity can be significantly reduced. As wave height increases, velocity increases, as shown in Figures 9a and 9b. A rectangular layout with VCD =12 had velocities of 1.34 m/s, 1.38 m/s, and 1.42 m/s for H=6, H=9, and H=12 cm, respectively. The triangular layout significantly decreased the total sediment rate by increasing drag force and cover resistance.

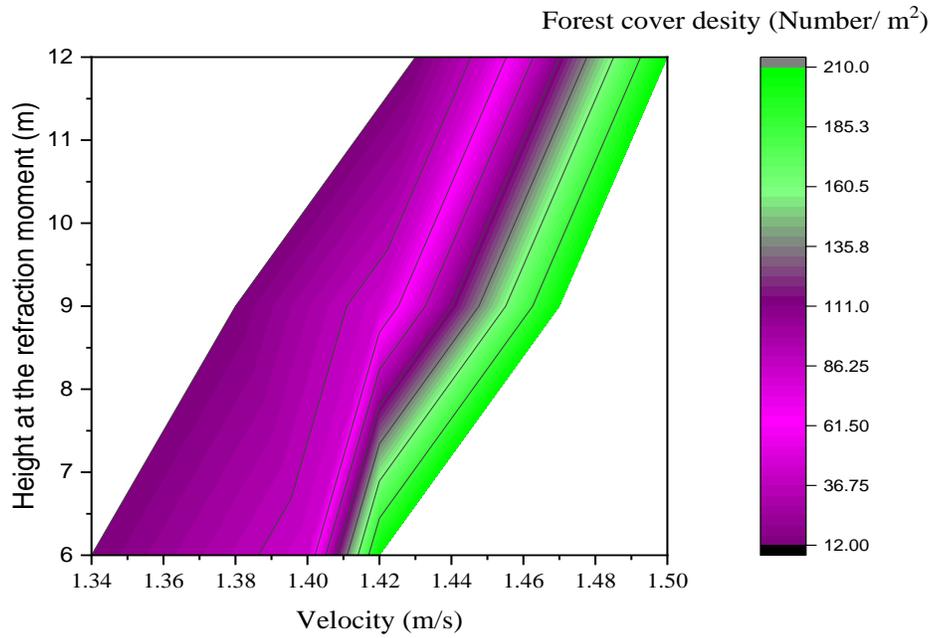

a

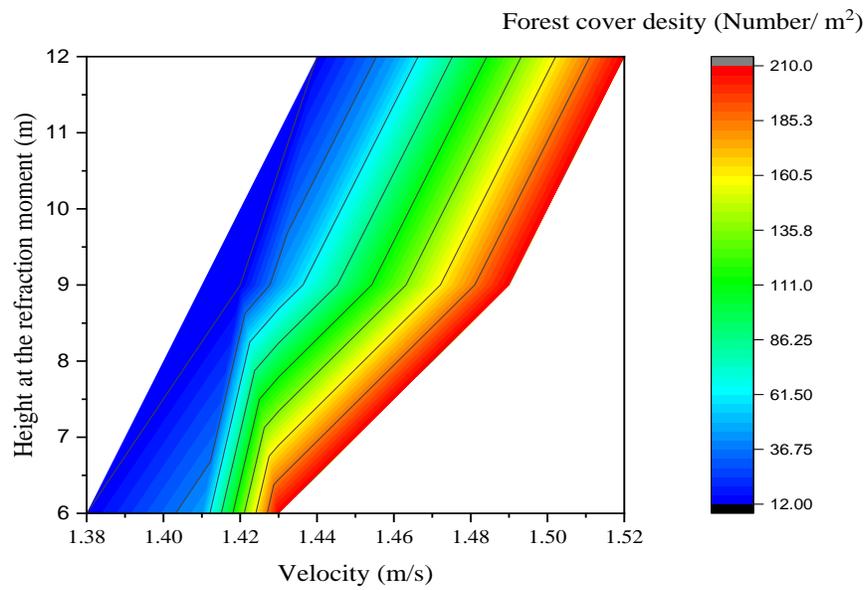

b

Figure 9. The investigation of height and wave velocity for different CFCs based on a: rectangular layout and b: triangular layout

In the R1 configuration, the dimensionless sediment transport rate (DSTR) is plotted against the dimensionless wave height ratio (H (at the refraction moment)/Y (sill height). DSTR is calculated by dividing the mass of sediments in the presence of vegetation cover by the mass of sediments without cover vegetation. Sediment transport rate increases with increasing wave height. When the number of rows of vegetation cover was equal to four, the DSTR for H/Y=0.65, H/Y=0.9, and H/Y=1.2 was 0.52, 0.58, and 0.61, respectively. An NRVC value represents how many rows of vegetation cover a particular area. Figures 10b, 10c, and 10d show DSTR versus H/Y plots for configurations R2, R3, and R4. In these figures, DSTR decreased as forest cover increased. In the H/Y=0.65 configuration of R4, the DSTR of NRVC=12, NRVC=10, NRVC=7, and NRVC=4 was 0.14, 0.20, 0.30, and 0.39, respectively. For the T1 configuration, Figure 10e shows DSTR versus H/Y.

An increase in H/Y increased DSTR. Considering four rows of vegetation cover and H/Y = 0.65, 0.9, and 1.2, the DSTR was 0.42, 0.50, and 0.55, respectively. Figures 10f, 10g, and 10 h show the DSTR versus H/Y for T2, T3, and T4 configurations. In the T4 configuration with H/Y=0.65, the DSTR of NRVC=12, NRVC=10, NRVC=7, and NRVC=4 were 0.12, 0.20, 0.25, and 0.36, respectively. A comparison of R1, R2, R3, and R4 configurations shows the R4 configuration produces less sediment.

The DSTRs of the R1, R2, R3, and R4 configurations were 0.53, 0.38, 0.20, and 0.12, respectively, in the H/Y=0.65 with NRVC=4. This implies that forest cover density increases as longitudinal and transverse distances decrease. When the vegetation cover density increases, the sedimentation rate can be significantly reduced.

Comparing T1, T2, T3, and T4 revealed that T4 has a higher sediment reduction rate. For T1, T2, T3, and T4 configurations, the DSTRs were 0.42, 0.28, 0.25, and 0.11 in the H/Y=0.65 with

NRVC=4. The triangular layout performed better than the rectangular layout in reducing DSTR. Under the NRVC=5, the DSTRs for the H/Y=0.65, H/Y=0.90, and H/Y=1.2 in the rectangular layout were 0.39, 0.41, and 0.42. Under NRVC=5, the DSTRs for the H/Y=0.65, H/Y=0.90, and H/Y=1.2 in the triangular layout were 0.28, 0.37, and 0.41.

The sediment transport rate versus coastal forest cover density is shown in figure 10i. Increasing density led to a significant decrease in sediment transport rate. Sediment transport rates ranged from 14 to 141 cm3/s. As forest cover density increases, sediment transport rate decreases by 90%. This study confirmed previous research findings. Jalil-Masir et al. (2021b) found that triangular layouts performed better than rectangular ones. Furthermore, Parnak et al. (2018) found that vegetation covers reduced sediment transport by 70%.

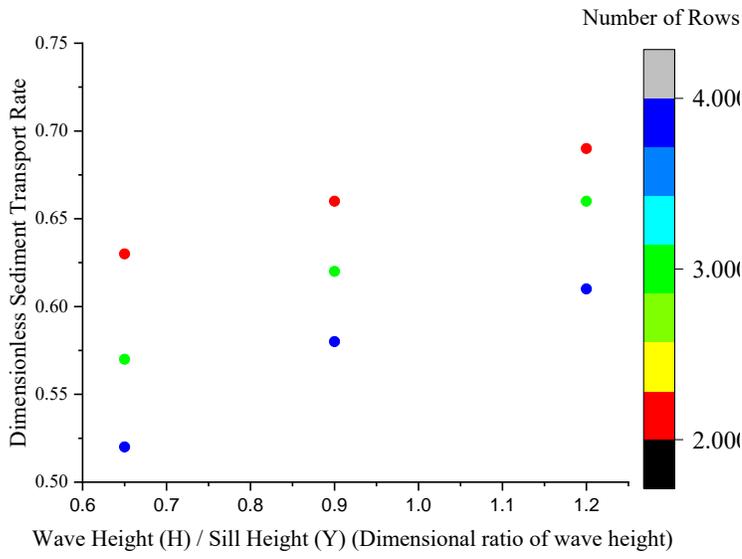

(a) (number of rows:2,3 ,and 4)

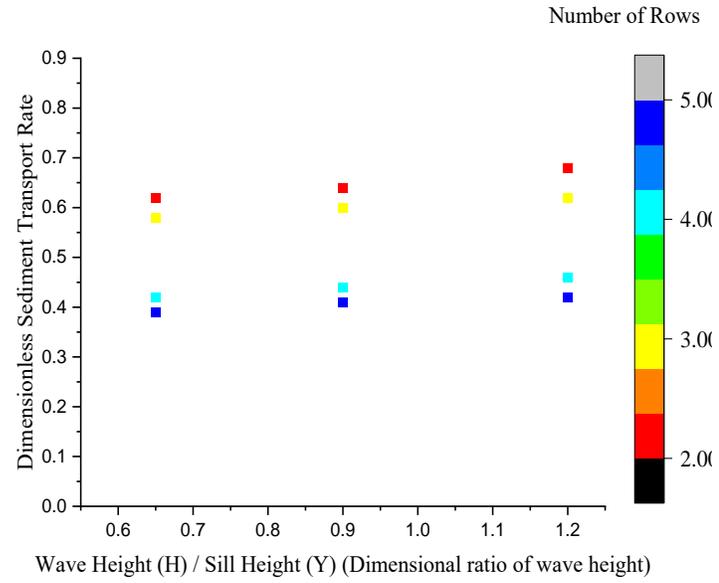

b (number of rows:2,3, 4, and 5)

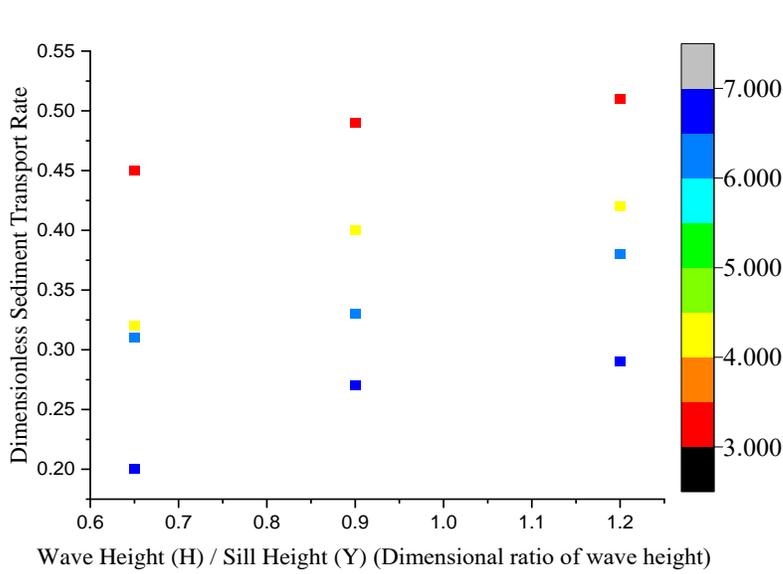

(c) (number of rows:7,6, 4, and 3)

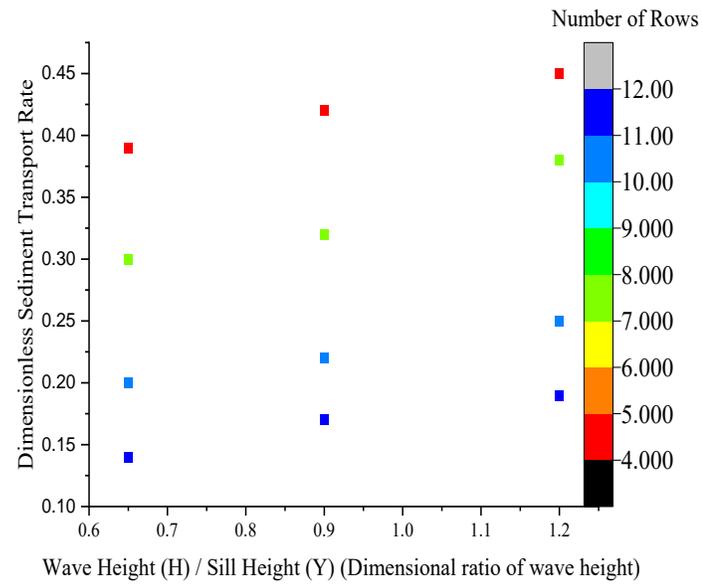

(d) (number of rows:13, 10, 7, and 4)

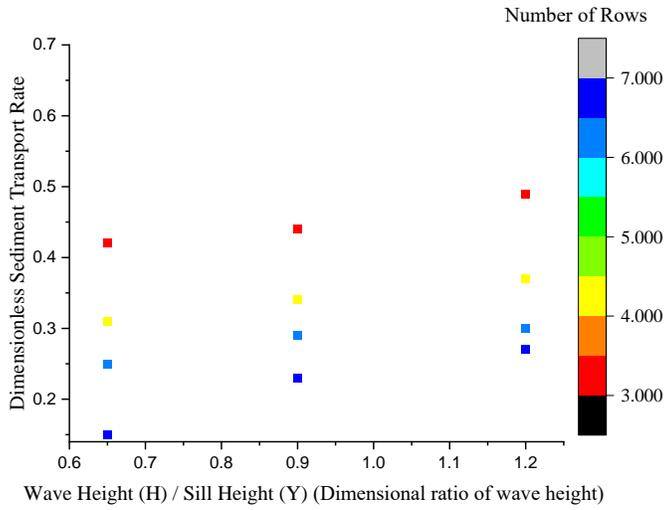

(e) (number of rows:7,6, 4, and 3)

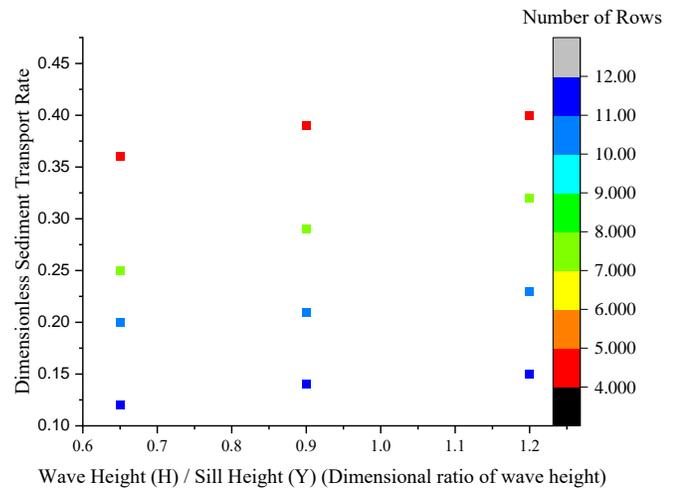

(f) (number of rows:13, 10, 7, and 4)

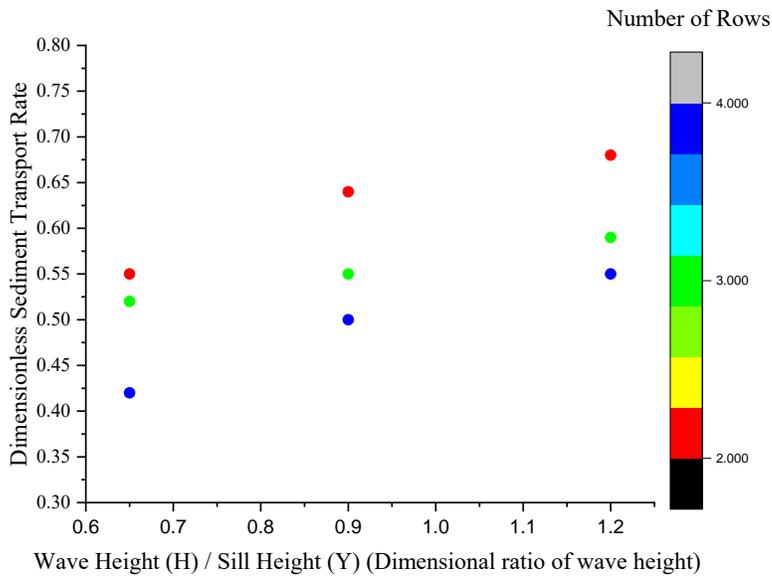

(g) (number of rows:2,3 ,and 4)

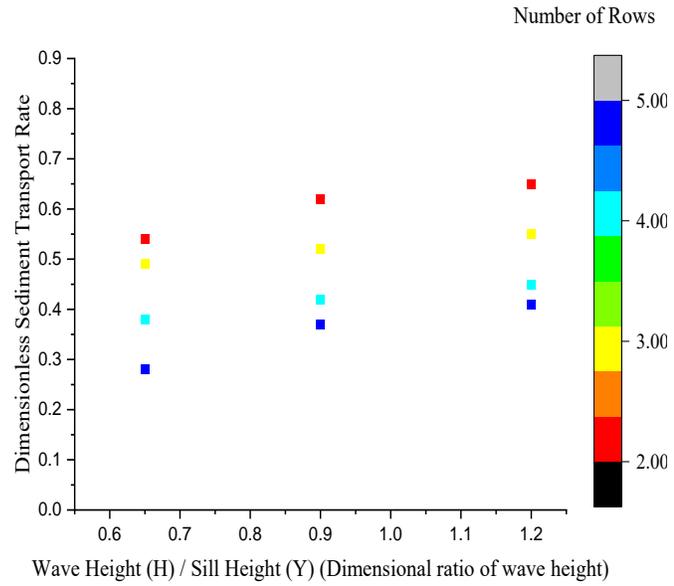

(h) (number of rows:2,3, 4, and 5)

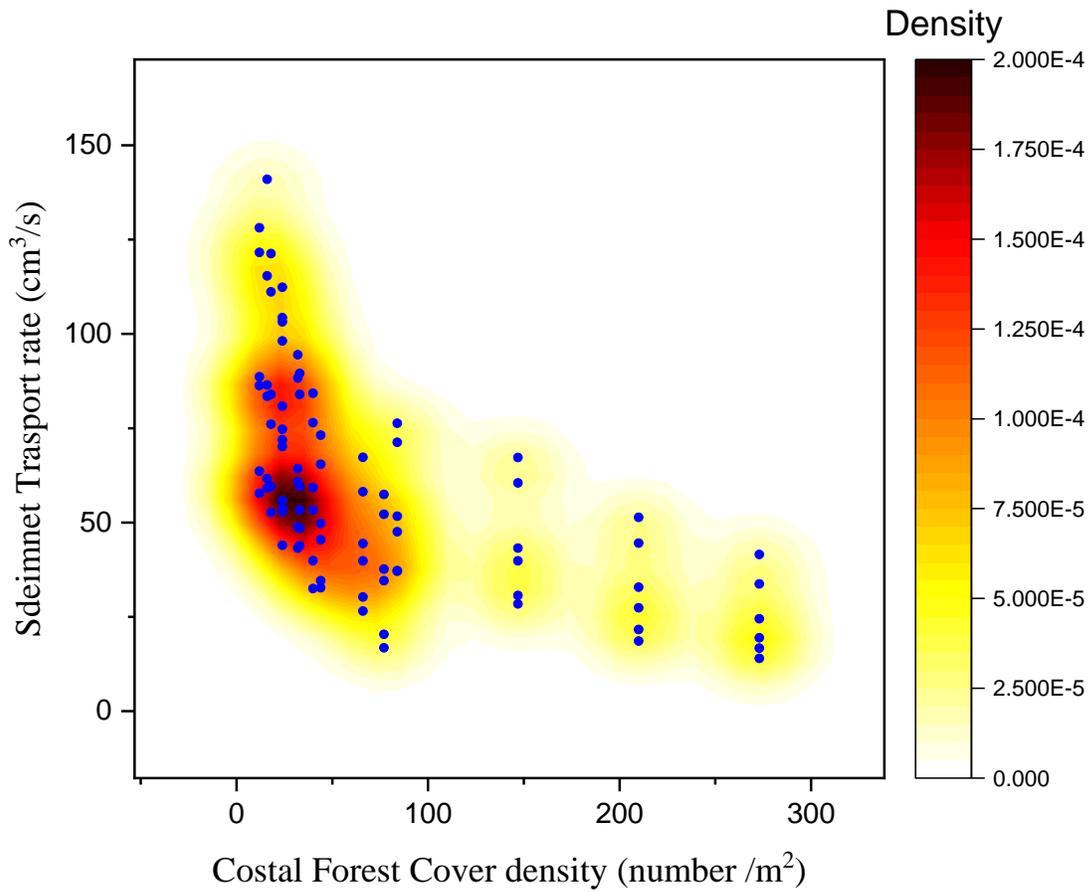

(i)

Figure 10. (a-d) The variation of sediment transport rate versus height wave for a: $R_1$ configuration, b: $R_2$ configuration, c: $R_3$ configuration, d: $R_4$ configuration

(e-h). The variation of sediment transport rate versus dimensionless wave height ratio in triangular layout for e: $T_1$ configuration, f: $T_2$ configuration, g: $T_3$ configuration, h: $T_4$ configuration

(i) (Variation of sediment transport rate versus the cover density)

From the obtained results, the following points infer:

1) Different vegetation layouts with triangular and rectangular layouts have different effects on sediment transfer rate (SDR).

2) The triangular layout is more efficient than the rectangular one due to its larger contact surface area.

3) Predicting sediment transfer rate requires determining input model parameters. Wave height and velocity significantly influence sediment transfer rate (STR) capacity.

4) The layout of vegetation is not limited to rectangular and triangular shapes. Future studies can also investigate other types of vegetation layouts, such as zigzag and random layouts.

5) Using outputs of several individual models in an ensemble model will remarkably strengthen the results' accuracy.

6) GMDH results were greatly improved by combining the GMDH model with optimized algorithms.

## 7) Conclusion

Predicting sediment transport rates is imperative for reducing the environmental pollution in watersheds. In coastal regions, robust models are essential for predicting sediment transport rates. Sediment transport can be reduced by planting coastal forests. This study investigated the effect of coastal forests on sediment transport through a comprehensive experiment. Various soft computing models were also used to predict sediment transport rates. The present study introduced a new ensemble model to predict sediment transport, and new optimization algorithms were developed to train GMDH models.

Inputs included sediment diameter, stem diameter, cover density, wave height, wave velocity, cover height, and wave force. The GMDH models were improved using the HBA, RSOA, SCA, and PSOA. Based on the outputs of GMDH models, an IMM model was used to predict the sediment transport rate. By combining several GMDH models, this model could take advantage of their strengths. The MAE of the IMM was 0.145 m3/s, while the MAEs of the GMDH-HBA, GMDH-RSOA, GMDH-SCA, GMDH-PSOA, and GMDH in the testing level were 0.176 $m^3$/s, 0.312 $m^3$/s, 0.367 $m^3$/s, 0.498 $m^3$/s, and 0.612 $m^3$/s, respectively. GMDH-HBA, GMDH-RSOA, GMDH-SCA, GMDH-PSOA, and GMDH had NSEs of 0.95, 0.93, 0.89, 0.86, 0.82, and 0.76, respectively.

It was found that coastal forecast cover was essential for reducing wave height and sediment transport rates due to its effect on sediment transport rates. A variety of configurations were used in the experiment. Triangular layouts reduced sediment transport more effectively than rectangular layouts. Various artificial neural networks and optimization algorithms can also forecast sediment transport. Additionally, the following studies can consider the impact of input uncertainty on sediment transport prediction.

All authors have agreed with the content and all have given explicit consent to publish.

**Competing interests**

The authors declare no competing interests.

**Availability of data and materials**

The datasets generated during and/or analyzed during the current study are available from the corresponding author on reasonable request.

**Funding**

This study was funded by the University of Shahrekord, Iran. The financial support of this organization is appreciated (GN: 141/5328).

**Acknowledgments**

This study was funded by the University of Shahrekord, Iran. The financial support of this organization is appreciated (GN: 141/3107).

**Competing interests**

The authors declare no competing interests.